\documentclass[twocolumn]{aastex63}

\usepackage{soul}
\usepackage{array}
\newcolumntype{S}{>{\arraybackslash}p{18cm}}

\def\apjl{ApJL}
\def\aap{A\&A}

\def\apjs{ApJ Supp}

\newcommand{\lsim }{{\lower0.8ex\hbox{$\buildrel <\over\sim$}}}
\newcommand{\gsim }{{\lower0.8ex\hbox{$\buildrel >\over\sim$}}}
\newcommand{\Msun}{\ifmmode {M_{\odot}}\else${M_{\odot}}$\fi}
\newcommand{\Lsun}{\ifmmode {L_{\odot}}\else${L_{\odot}}$\fi}
\newcommand{\Rsun}{\ifmmode {R_{\odot}}\else${R_{\odot}}$\fi}
\newcommand{\ergs}{erg~s$^{-1}$}
\newcommand{\ergcms}{erg~s$^{-1}$~cm$^{-2}$}

\newcommand{\chandra}{\textit{Chandra}}
\newcommand{\acis}{\textit{Chandra}/ACIS}
\newcommand{\hrc}{\textit{Chandra}/HRC}

\newcommand{\integral}{\textit{INTEGRAL}}
\newcommand{\rosat}{\textit{ROSAT}}

\shorttitle{XRBs in Globular Clusters}
\shortauthors{Bahramian et al.}
\graphicspath{{./}{figures/}}

\begin{document}

\title{The MAVERIC Survey: \acis\ Catalog of Faint X-ray sources in 38 Galactic globular clusters}

\correspondingauthor{Arash Bahramian}
\email{arash.bahramian@curtin.edu.au}

\author[0000-0003-2506-6041]{Arash Bahramian}
\affil{International Centre for Radio Astronomy Research – Curtin University, GPO Box U1987, Perth, WA 6845, Australia}

\author[0000-0002-1468-9668]{Jay Strader}
\affil{Center for Data Intensive and Time Domain Astronomy, Department of Physics and Astronomy, Michigan State University, East Lansing, Michigan 48824, USA}

\author[0000-0003-3124-2814]{James C. A. Miller-Jones}
\affil{International Centre for Radio Astronomy Research – Curtin University, GPO Box U1987, Perth, WA 6845, Australia}

\author[0000-0002-8400-3705]{Laura Chomiuk}
\affil{Center for Data Intensive and Time Domain Astronomy, Department of Physics and Astronomy, Michigan State University, East Lansing, Michigan 48824, USA}

\author[0000-0003-3944-6109]{Craig O. Heinke}
\affil{Physics Department, University of Alberta, 4-183 CCIS, Edmonton, AB T6G 2E1, Canada}

\author[0000-0003-0976-4755]{Thomas J. Maccarone}
\affil{Department of Physics and Astronomy, Texas Tech University, Box 41051, Lubbock, TX 79409–1051, USA}

\author[0000-0003-4897-7833]{David Pooley}
\affil{Department of Physics and Astronomy, Trinity University, San Antonio, TX, USA}

\author{Laura Shishkovsky}
\affil{Center for Data Intensive and Time Domain Astronomy, Department of Physics and Astronomy, Michigan State University, East Lansing, Michigan 48824, USA}

\author[0000-0003-4553-4607]{Vlad Tudor}
\affil{International Centre for Radio Astronomy Research – Curtin University, GPO Box U1987, Perth, WA 6845, Australia}

\author[0000-0002-9547-8677]{Yue Zhao}
\affil{Physics Department, University of Alberta, 4-183 CCIS, Edmonton, AB T6G 2E1, Canada}

\author[0000-0002-0439-7047]{Kwan Lok Li}
\affil{Institute of Astronomy, National Tsing Hua University, Hsinchu 30013, Taiwan}

\author[0000-0001-6682-916X]{Gregory R. Sivakoff}
\affil{Physics Department, University of Alberta, 4-183 CCIS, Edmonton, AB T6G 2E1, Canada}

\author[0000-0002-4039-6703]{Evangelia Tremou}
\affil{LESIA, Observatoire de Paris, CNRS, PSL, SU/UPD, Meudon, France}

\author[0000-0003-0426-6634]{Johannes Buchner}
\affil{Max-Planck-Institut f\"{u}r extraterrestrische Physik, Gie\ss enbachstra\ss e 1, 85748 Garching, Germany}



\begin{abstract}

Globular clusters host a variety of lower-luminosity ($L_X<10^{35}$ \ergs) X-ray sources, including accreting neutron stars and black holes, millisecond pulsars, cataclysmic variables, and chromospherically active binaries. In this paper, we provide a comprehensive catalog of more than 1100 X-ray sources in 38 Galactic globular clusters observed by the \chandra\ X-ray Observatory's \acis\ detector. The targets are selected to complement the MAVERIC survey's deep radio continuum maps of Galactic globular clusters. We perform photometry and spectral analysis for each source, determine a best-fit model, and assess the possibility of it being a foreground/background source based on its spectral properties and location in the cluster. We also provide basic assessments of variability. We discuss the distribution of X-ray binaries in globular clusters, their X-ray luminosity function, and carefully analyze  systems with $L_X > 10^{33}$ \ergs. Among these moderately bright systems, we discover a new source in NGC~6539 that may be a candidate accreting stellar-mass black hole or a transitional millisecond pulsar. We show that quiescent neutron star LMXBs in globular clusters may spend $\sim$2\% of their lifetimes as transitional millisecond pulsars in their active ($L_X>10^{33}$ \ergs) state. Finally, we identify a substantial under-abundance of bright ($L_X>10^{33}$ \ergs) intermediate polars in globular clusters compared to the Galactic field, in contrast with the literature of the past two decades.

\end{abstract}



\section{Introduction}
An X-ray binary (XRB) is a system where a compact object (white dwarf, neutron star, or black hole) is accreting from a companion star. The companion is typically a main sequence star, but can also be an evolved star or a white dwarf. Observations have shown that XRBs are significantly overabundant in globular clusters (GCs) compared to the field \citep[both in the Milky Way and other galaxies,][]{Clark75,Sarazin03,Kundu07a, Kundu07b}. This overabundance has been linked to dynamical XRB formation channels in GCs. In the field, these systems form through binary evolution. In GCs, many are formed through stellar interactions made possible due to the high population density in GCs. These channels include interaction of a red giant with a compact object \citep{Sutantyo75}, tidal capture of a companion by the compact object \citep{Fabian75}, or an exchange interaction, where a compact object replaces a low mass star in a binary \citep{Hills76}.

These formation channels rely strongly on encounters in dense environments, and thus they are expected to correlate with some properties of GCs. One cluster property that influences the frequency of encounters---and thus the production of XRBs---is the ``stellar encounter rate'', generally written as $\Gamma\propto\int\frac{\rho^2}{\sigma}dV$, integrated over the GC volume, where $\rho$ is the cluster density and $\sigma$ is the cluster velocity dispersion \citep{verbunt87}. A strong correlation between the population of XRBs and stellar encounter rate has been observed in Galactic GCs \citep{Pooley03, Heinke03d, Bahramian13}, confirming that XRBs are formed through encounters in dense clusters. Dynamical interactions can also destroy primordial binaries (e.g., \citealt{Davies97}), and this process can dominate over dynamical formation in lower-density GCs \citep{Heinke20}.

X-ray sources in GCs consist of multiple sub-populations. These include XRBs like cataclysmic variables (CVs; where a white dwarf [WD] accretes from a low mass star), low-mass X-ray binaries (LMXBs) with a black hole (BH) or neutron star (NS) accreting from a low-mass ($\lsim M_\odot$) star, and millisecond pulsars (MSPs), which are thought to be descendants of NS-LMXBs. Furthermore, there is also a small group of bright chromospherically active binaries (ABs), which contain two tidally locked main sequence stars in close orbit. The tidally locked orbit in these systems allows their combined X-ray luminosity to reach $\gsim 10^{30}$ \ergs, likely through an enhanced dynamo resulting from the increased rotational velocities (e.g., \citealt{Walter81}). While ABs are abundant in GCs and form a large fraction of sources with $\lsim 10^{31}$ \ergs, only a small fraction of them reach $\gsim 10^{31}$ \ergs. It has been demonstrated that fainter X-ray sources in GCs (which are mostly ABs) are likely primordial in origin as their numbers appear correlated with GC mass \citep{Bassa04, Bassa08, Cheng18, Heinke20}.

It is difficult to confidently determine the nature of a faint XRB in a GC based on X-rays alone, and objects thought to be confidently classified primarily from X-rays later turn out to have been typed incorrectly \citep[e.g.,][]{Miller-Jones15}. Adding in observations from additional wavelengths has helped in classifying numerous X-ray sources in various Galactic GCs \citep[e.g.,][]{Grindlay01a, Pooley02a, Edmonds03b, Heinke06b, Lugger07, Maxwell12}. Still, the nature of many GC X-ray sources remains elusive. 

CVs are the most abundant group of XRBs in GCs \citep[e.g.][]{Grindlay01a, Pooley02a,Cohn10,Lugger17,RiveraSandoval18b}. CVs in GCs can be formed primordially (via binary formation and evolution), or through dynamical interactions \citep{Davies97,Ivanova06, Shara06, Hong17, Belloni19}. However, most present-day CVs in GCs appear to be formed dynamically \citep{Ivanova06}, and it has been suggested that most of the primordial CVs have been disrupted \citep[at least in some GCs;][]{Davies97}.

Almost all bright ($L_X > 10^{34}$ \ergs) XRBs in Galactic GCs have been proven to be NS-LMXBs\footnote{\url{https://bersavosh.github.io/research/gc_lmxbs.html}} \citep[e.g.,][]{Bahramian14}, either through detection of pulsations or X-ray bursts. Additionally, numerous faint ($L_X < 10^{34}$ \ergs) X-ray sources in Galactic GCs have been identified as quiescent NS-LMXBs, typically through the detection of a thermal component consistent with a neutron star \citep[e.g.,][]{Grindlay01a,Rutledge02a, Pooley02a, Heinke03a, Heinke05a, Servillat08a, Guillot09b, Maxwell12}. The quiescent NS-LMXBs in GCs (with low reddening) provide some of the best samples to study properties of NSs like low-level accretion, crustal cooling, and constraints on mass and radius of NSs \citep{Degenaar11b, Bahramian15, RiveraSandoval18a, Ootes19, Guillot13, Steiner18}.

GCs harbour a large population of radio millisecond pulsars \citep{Ransom05,Hessels07,Freire10,Dai20}, and some of which have been detected in X-rays \citep[e.g.][]{Bhattacharya17, Bogdanov10b}. MSPs are the ``recycled'' descendants of NS-LMXBs. This link was first proposed upon discovery of the first rotation-powered millisecond pulsar \citep{Alpar82} and the observation of accreting millisecond X-ray pulsars provided strong indirect evidence in support of it \citep{Wijnands98}. It was strengthened further by the discovery of transitional millisecond pulsars (tMSPs), which are NS-LMXBs that switch between rotation-powered radio pulsations and accretion-powered X-ray pulsations \citep{Archibald09, Papitto13, Linares14b, deMartino10, deMartino13, Bassa14}. Today, there are $>150$ MSPs identified from radio observations of GCs\footnote{See P. Freire's catalog: \url{http://naic.edu/~pfreire/GCpsr.html}} showing a link between dynamical formation and their population \citep{Bagchi11, Hui11, Bahramian13}. Additionally, while there are still very few tMSPs known to date, at least two (one confirmed and one candidate) are located in GCs, hinting at a possible overabundance of tMSPs in GCs \citep[IGR J18245--2452 in M~28 and Terzan~5 CX1 in Terzan~5;][]{Papitto13, Bahramian18}.

The presence of black holes (BHs) in GCs has been heavily debated over the past few decades. While initial stellar evolution in GCs is expected to have produced a high number of BHs, there are very few confirmed in them so far. Theoretical studies once favored the contention that all (or almost all) BHs would mass-segregate to the cluster core and get kicked out of GCs rather quickly \citep{Sigurdsson93}. This contention appears to be consistent with observational studies of bright ($L_X > 10^{35}$ \ergs) XRBs in GCs, which showed that all of these systems are NS-LMXBs \citep{Verbunt06, Papitto13, Bahramian14}. This strengthened theories that perhaps {\it all} BHs get kicked out of GCs \citep{Kulkarni93}. In recent years, detection of accreting BH candidates in Galactic and extra-galactic GCs \citep[e.g.,][]{ Maccarone07a,Strader12, Chomiuk13, Miller-Jones15, Shishkovsky18}, and discovery of dynamically-confirmed detached BHs in wide-orbit binaries in the Galactic GC NGC~3201 \citep{Giesers18,Giesers19}, reversed these earlier suggestions, and indicated that a subset of BHs do indeed survive in GCs. A parallel revolution in theoretical work now suggests that many BHs are likely to be retained in GCs \citep[e.g.,][]{Morscher13, Morscher15}, and perhaps even form BH-BH binaries that merge and can be detected by gravitational wave observatories \citep{Rodriguez16}.

Determining the nature of the compact object in LMXBs is challenging. In bright outbursting or persistent systems, the detection of pulsations or X-ray bursts indicates an NS accretor, but the absence of these signatures does not reject it. The problem becomes more difficult in quiescent systems where these signatures are challenging or impossible to detect. For some binaries, components in the X-ray spectrum, like a soft blackbody-like emission from the surface of the neutron star in NS-LMXBs, can help determine the nature of the source \citep[e.g.,][]{Rutledge02a}. Another method, which has come into vogue in the last decade, is based on the  coupling between the accretion flow and the possible presence of a radio jet: while both BHs and NSs show evidence for jets, BHs tend to be brighter in the radio than NSs by a factor of $\sim$5 to 20 (at least at some X-ray luminosities; \citealt{Fender03, Migliari06, Tudor17, Gallo18}).

The MAVERIC survey is focused on identifying accreting BH candidates and other exotic binaries in a large sample of Galactic GCs through radio observations with the Karl G.\ Jansky Very Large Array (VLA) and the Australia Telescope Compact Array (ATCA) \citep{Shishkovsky18, Tremou18}. The goal of the present work is to produce a catalog of X-ray sources in all Galactic GCs without existing comprehensive X-ray studies to accompany the MAVERIC radio continuum data. 

In this paper, we provide a full catalog of faint X-ray sources in 38 Galactic GCs based on analysis of available \acis\ observations. In \S\ref{sec:data}, we present sample selection, data, and reduction; in \S\ref{sec:analysis}, we provide details of our analysis method, and in \S\ref{sec:catalog}, we present the source catalog and discuss various measured and estimated source properties in the catalog. Finally in \S\ref{sec:disc}, we discuss the impact of this catalog on our understanding of the XRB population in GCs and the nature of the brightest sources in this catalog, including a new BH-LMXB candidate. This study, along with the MAVERIC survey (Shishkovsky et al.\ in prep., Tudor et al.\ in prep.), will enable a careful multi-wavelength study of accreting systems in Galactic GCs. 

\section{Data and Reduction}\label{sec:data}
\subsection{Sample selection}\label{sec:sample}
The MAVERIC sample was initially chosen as all Galactic GCs within 9 kpc and $> 10^5 M_{\odot}$, plus a few more massive GCs at larger distances, for a total sample of 50 GCs. The present catalog only includes GCs that have been observed by \acis\ at least once and that have no X-ray sources in outburst within the cluster. This selection ensures the presence of spectral information from \acis\ (as opposed to \hrc, which lacks sufficient spectral resolution for our purposes). 

The extended tail of the point-spread function (PSF) for bright ($F_X\gg10^{-12}$ \ergcms) X-ray sources severely impacts the detection threshold and completeness for nearby faint sources in \acis\ observations. Thus some GCs are excluded due to persistent bright XRBs (M~15, NGC~6441, NGC~6624, NGC~6652, NGC~6712), while others (Liller~1, Terzan~6) had \chandra\ observations only taken during bright outbursts. 

We also excluded Omega~Cen, 47~Tuc, and NGC~6752 as deep X-ray source catalogs on these clusters have been recently published \citep{Henleywillis18, Bhattacharya17, Cheng19a, Forestell14}. 

These criteria provide us with a sample of 38 GCs (Table~\ref{tab:obslist}). Of these 38, 10 have new data taken as part of MAVERIC: 6 (Djorg 2, M~10, M~19, NGC~4372, NGC~4833, and M~107) have single \acis\ observations obtained specifically for MAVERIC follow-up, while 4 others (M~22, M~62, Terzan~1, and Terzan~5) have both previous data and newer MAVERIC follow-up data.

\subsection{Data Reduction}
We used \textsc{CIAO} 4.8 and higher (as there were software updates over the duration of the project), with CalDB 4.7.7 and higher for all \chandra\ data reduction\footnote{We note that there have been no software/calibration updates that would impact the analysis in this work significantly.} \citep{Fruscione06}, extracted data products and performed analysis using \textsc{ACIS-Extract} \citep[here after AE,][]{Broos10} and Bayesian X-ray Analysis package \textsc{BXA} \citep{Buchner14}. Our overall goal was to obtain a carefully vetted, consistently analyzed X-ray source catalog for the GCs in our sample. Thus, we first reprocessed the data and produced input files necessary for a high-level analysis (event files, exposure maps, aspect solutions, source catalogs etc) using \textsc{CIAO} and used these as input (after inspection and modification, where necessary) for analysis in AE. Finally we used the spectra produced by AE for spectral analysis in \textsc{BXA} (\S\ref{sec:analysis}). We checked all observations for background flares. We found noticeable flares in four observations out of 90 analyzed in this work. Out of these four, three are negligible due to their short duration compared to the full observation and weak amplitude (Obs.IDs 2683, 3798, 18997 in M~28, Terzan~5, and M~30, respectively). However, flaring in one observation (Obs. ID 5435 in NGC~6544) is strong (increasing the background by a factor of $\sim$8) and lasts for 5 ks. Thus for the case of obs.ID 5435, we followed the standard procedure to remove background flares.\footnote{\url{https://cxc.cfa.harvard.edu/ciao/threads/flare/index.html}}

Given the large amount of data, we partially automated the data reduction process using the module \texttt{ciao\_contrib.runtool} in \textsc{Python}, provided in the \textsc{CIAO} package. All observations were first reprocessed using \texttt{chandra\_repro}, then we produced exposure maps and image files using the task \texttt{fluximage} in the 0.5--10 keV band, with an image binning factor of 0.5. A small ($< 1$) binning factor can improve detection of faint sources in crowded regions (like the core of some GCs) on the \acis\ detector.

\begin{figure}
\centering
\includegraphics[scale=0.42]{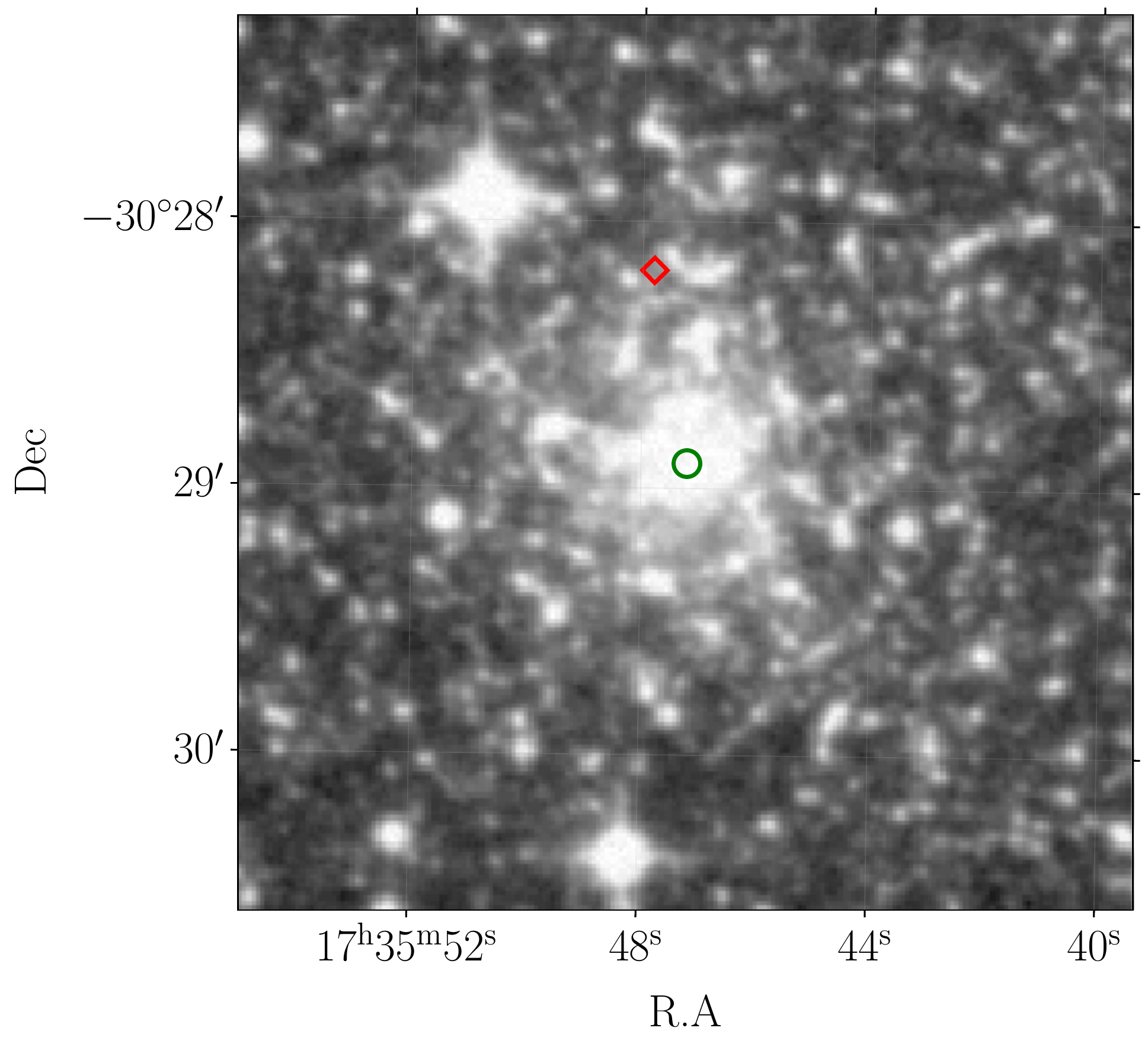}
\caption{DSS image of Terzan~1 and comparison of GC coordinates from the Harris catalog (red diamond) to the values from \citet[][green circle]{Picard95}.}
\label{fig:ter1}
\end{figure}

In many observations covering the GCs in our sample, the fields of view (FOV) covered are substantially larger than the half-light radius of the target GC. As we are only interested in sources within the GCs, we filtered all images, exposure maps and event files using \texttt{dmcopy} to only include the data covering the GC up to 1.2 times the half-light radius. The  factor of 1.2 is a rather conservative upper limit to consider uncertainties in measurements of half-light radii and GC center coordinates. We used GC center coordinates and half-light radii from the Harris catalog \citep[2010 edition,][]{Harris96} to apply this cut. We note that the Harris catalog coordinates for Terzan~1 appear to have a $\sim44''$ offset with the values reported by \citet{Picard95}. When compared to images of the cluster, the values from \citet{Picard95} appear to be more accurate (Figure~\ref{fig:ter1}). Thus, for this specific cluster, we used the cluster coordinates from \citet{Picard95}.

After restricting the observation and analysis to the region of interest, we ran \texttt{wavdetect} on each observation to produce observation-specific source catalogs (source detection is discussed in detail in \S\ref{sec:detection}). If there were more than one observation covering a GC, we used \texttt{merge\_obs} to reproject and merge all available data on the GC. For most GCs, comparing observation-specific source catalogs indicated that the relative astrometric shifts between observations of a GC are negligible. However, this was not the case for Terzan~5 and M~30. For these clusters, we used \texttt{wcs\_match} and \texttt{wcs\_update} to correct the relative astrometry in the event files and aspect solutions, prior to producing images and merging observations.

\begin{deluxetable*}{llcccllcc}
\tablecaption{List of all GCs and their \acis\ observations reduced and analyzed in this work.}
\label{tab:obslist}
\tablehead{
\colhead{GC}      &   \colhead{Obs.ID}  & \colhead{Date}      & \colhead{Exp (ks)} & \vline & \colhead{GC}      &   \colhead{Obs.ID}  & \colhead{Date}      & \colhead{Exp (ks)}
}
\startdata
Djorg~2	& 17844           & 2017-05-13 & 22.7  &\vline  & NGC~5927  & 08953           & 2008-04-30 & 7.8  \\
M~2	    & 08960           & 2008-04-09 & 11.5  &\vline  & 		    & 13673           & 2012-01-03 & 47.4 \\
M~4	    & 00946           & 2000-06-30 & 25.8  &\vline  & NGC~6139  & 08965           & 2008-05-26 & 17.7 \\
		& 07446           & 2007-09-18 & 47.9  &\vline  & NGC~6304  & 08952           & 2008-01-28 & 5.3  \\
		& 07447           & 2007-07-06 & 45.5  &\vline  &    	    & 11073$^*$       & 2010-07-06 & 97.5 \\
M~5	    & 02676           & 2002-09-24 & 44.7  &\vline  & NGC~6325  & 08959           & 2009-01-29 & 16.7 \\
M~9	    & 08954           & 2009-01-29 & 8.4   &\vline  & NGC~6352  & 13674           & 2012-06-20 & 19.8 \\
M~10    & 16714           & 2015-05-08 & 32.6  &\vline  & NGC~6362  & 11024           & 2009-11-21 & 30.8 \\
M~12	& 04530           & 2004-07-17 & 26.6  &\vline  &   	    & 12038           & 2009-11-29 & 9.0  \\
M~13	& 05436           & 2006-03-11 & 26.8  &\vline  & NGC~6388  & 05505           & 2005-04-21 & 44.6 \\
		& 07290           & 2006-03-09 & 27.9  &\vline  &  		    & 12453$^\dagger$ & 2011-08-29 & 2.5  \\
M~14	& 08947           & 2008-05-24 & 12.1  &\vline  & NGC~6397  & 00079$^*$       & 2000-07-31 & 48.3 \\
M~19	& 17848           & 2016-07-02 & 22.7  &\vline  &    	    & 02668           & 2002-05-13 & 28.1 \\
M~22	& 05437           & 2005-05-24 & 15.8  &\vline  &    	    & 02669           & 2002-05-15 & 26.7 \\
		& 14609           & 2014-05-22 & 84.9  &\vline  &    	    & 07460           & 2007-07-16 & 147.7\\
M~28    & 02683           & 2002-09-09 & 14.1  &\vline  &    	    & 07461           & 2007-06-22 & 88.9 \\
		& 02684           & 2002-07-04 & 12.8  &\vline  & NGC~6522  & 03780$^*$       & 2003-07-09 & 96.5 \\
		& 02685           & 2002-08-04 & 13.5  &\vline  &    	    & 08963           & 2008-10-27 & 8.3  \\
		& 09132           & 2008-08-07 & 142.3 &\vline  & NGC~6539  & 08949           & 2008-10-31 & 15.0 \\
		& 09133           & 2008-08-10 & 54.5  &\vline  & NGC~6541  & 03779           & 2003-07-12 & 44.8 \\
		& 16748           & 2015-05-30 & 29.7  &\vline  & NGC~6544  & 05435           & 2005-07-20 & 16.3 \\
		& 16749           & 2015-08-07 & 29.6  &\vline  & NGC~6553  & 08957           & 2008-10-30 & 5.2  \\
		& 16750           & 2015-11-07 & 29.6  &\vline  &    	    & 13671           & 2013-03-15 & 31.5 \\
M~30	& 02679           & 2001-11-19 & 49.4  &\vline  & NGC~6760  & 13672           & 2012-06-27 & 51.4 \\
		& 18997$^\dagger$ & 2017-09-06 & 90.2  &\vline  & Terzan~1  & 05464           & 2005-05-10 & 18.9 \\
		& 20725$^\dagger$ & 2017-09-04 & 17.5  &\vline  &    	    & 17847           & 2017-05-11 & 14.9 \\
		& 20726$^\dagger$ & 2017-09-10 & 19.2  &\vline  &    	    & 20075           & 2017-05-10 & 13.8 \\
		& 20731$^\dagger$ & 2017-09-16 & 24.0  &\vline  & Terzan~5  & 03798           & 2003-07-13 & 39.3 \\
		& 20732$^\dagger$ & 2017-09-14 & 47.9  &\vline  &    	    & 10059$^\dagger$ & 2009-07-15 & 36.3 \\
		& 20792$^\dagger$ & 2017-09-18 & 36.9  &\vline  &    	    & 13225           & 2011-02-17 & 29.7 \\
		& 20795$^\dagger$ & 2017-09-22 & 14.3  &\vline  &    	    & 13252           & 2011-04-29 & 39.5 \\
		& 20796$^\dagger$ & 2017-09-23 & 30.7  &\vline  &    	    & 13705           & 2011-09-05 & 13.9 \\
M~54    & 04448$^*$       & 2003-09-01 & 29.8  &\vline  &    	    & 13706           & 2012-05-13 & 46.5 \\
M~55    & 04531           & 2004-05-11 & 33.7  &\vline  &    	    & 14339           & 2011-09-08 & 34.1 \\
M~62    & 02677           & 2002-05-12 & 62.3  &\vline  &    	    & 14475$^\dagger$ & 2012-09-17 & 30.5 \\
		& 15761$^\dagger$ & 2014-05-05 & 82.1  &\vline  &    	    & 14476$^\dagger$ & 2012-10-28 & 28.6 \\
M~92    & 03778           & 2003-10-05 & 29.7  &\vline  &    	    & 14477$^\dagger$ & 2013-02-05 & 28.6 \\
		& 05241           & 2003-10-19 & 22.9  &\vline  &    	    & 14478$^\dagger$ & 2013-07-16 & 28.6 \\
M~107	& 17845           & 2016-06-28 & 11.8  &\vline  &    	    & 14479$^\dagger$ & 2014-07-15 & 28.6 \\
NGC~2808& 07453$^*$       & 2007-06-19 & 45.4  &\vline  &    	    & 14625           & 2013-02-22 & 49.2 \\
		& 08560$^*$       & 2007-06-21 & 10.8  &\vline  &    	    & 15615           & 2013-02-23 & 84.2 \\
NGC~3201& 11031           & 2010-09-22 & 83.5  &\vline  &    	    & 15750           & 2014-07-20 & 23.0 \\
NGC~4372& 17843           & 2016-06-25 & 10.3  &\vline  &    	    & 16638           & 2014-07-17 & 71.6 \\
NGC~4833& 17846           & 2017-01-03 & 11.8  &\vline  &   	    & 17779           & 2016-07-13 & 68.8 \\
        &                 &            &       &\vline  &           & 18881           & 2016-07-15 & 64.7 \\
\enddata
\tablecomments{$*$ indicates that the observation was taken with the target on ACIS-I (as opposed to ACIS-S). $\dagger$ indicates the observation was taken in sub-array mode.}
\end{deluxetable*}

\section{Analysis Methods}\label{sec:analysis}
After reprocessing and preparing the \acis\ data on each cluster, we proceeded to perform source detection, followed by photometry and source evaluation in AE, and spectral analysis in AE and BXA. 

\subsection{Source Detection}\label{sec:detection}
We performed source detection on all observations/GCs by running \texttt{wavdetect} on the under-binned stacked images (binned by a factor of 0.5) of each cluster. Source detection was carried out with a broad range of scales (1, 2, 4, 8, 16), and generally liberal detection thresholds (\texttt{wavdetect}'s threshold for identifying a pixel as belonging to a source, $10^{-4} - 10^{-5}$, depending on the depth and noise level in the observations). Choosing a liberal detection threshold generally increases the number of possible false detections, while it allows sensitivity to marginal detections. We chose such a threshold in order to leave assessment of whether a source is real to a later stage in AE, which allows a more rigorous verification in the photon-starved regime.

Merging multiple data sets generally produces deeper data; however in some cases it produced sub-optimal results. For example, in one of the two \acis\ observations of M~62 (Obs.\ ID 02677), the cluster core is $\sim 2.4'$ off-axis, while it is on-axis in the other observation (Obs.\ ID 15761). Thus, merging these two observations produced artifacts caused by the different PSF. This is a particularly important issue for M~62, as the cluster contains numerous X-ray sources in its core. This issue is also present in observations covering NGC~6522, where the first observation (Obs.\ ID 03780) targets Baade's window, but also covers this cluster at a highly off-axis angle. In these cases, instead of producing a source catalog based on the merged image, we cross-matched source catalogs from the two observations to produce a complete source catalog. 

Another issue impacting some GCs was the presence of faint outbursts. While we excluded observations in which a bright outburst (or persistent emission) from an XRB prevented effective inspection of faint sources in the cluster, we did include observations where outbursts were sufficiently faint to allow detection of faint sources in the GC to some level. There were only two observations impacted by this issue (one for Terzan~5, and one for NGC~6388). In both cases, the impact of the contaminated observations was small, particularly as the rest of the data covering these clusters were relatively deeper.

After producing observation-specific and merged source catalogs for each GC, we visually inspected the images and detected sources to evaluate the quality of the stacked images and search for clear sources that were missed by the algorithm, especially in crowded regions (where \texttt{wavdetect} has difficulty picking two sources close together apart). If there were obvious sources that were missed after fine-tuning the source detection parameters (e.g., detection threshold in \texttt{wavdetect}), we made a catalog of these sources manually for each cluster, with coordinates estimated based on rough centroid coordinates. These manually-added input catalogs were only produced for 6 GCs, and contain a total of 31 sources together. These consist of 10 possible sources in Terzan~5, 10 in M~62, 8 in M~28, and 1 in each of M~30, M~92, and NGC~6397. These catalogs provide an additional list of sources for verification and analysis in AE.

After producing filtered event files, aspect solutions, exposure maps through \textsc{CIAO}, and source catalogs as described above, we passed these as input to AE to evaluate source properties, extract light curves and spectra, and perform spectral analysis (with \textsc{BXA}).

\subsection{Source evaluation and photometry}
We used AE (version 2018feb6) to assess source significance, enhance source localization, extract source-specific products (e.g., event lists, light curves, spectra and associated files), and perform photometry. We followed the general recipe provided in the AE manual\footnote{\url{http://personal.psu.edu/psb6/TARA/ae\_users\_guide.html}} for point sources, with some modifications, as detailed below. 

Background extraction is a crucial component for most parts of our analysis (source validation, photometry and spectral analysis) and it requires additional care. AE provides multiple methods for selection of a background region based on complexity of emission and sources in the field. In most cases we used circular regions with model-based masking produced via \texttt{ae\_better\_masking} (selecting larger masks for brighter sources and smaller masks for faint sources).
However, in cases of crowded fields, this method can produce unusually large background regions with almost all local background ignored under masks. Thus for GCs where crowding is a possible issue (M~28, M~30, M~62, M~92, NGC~6388, NGC~6397, NGC~6541, and Terzan~5), we used model-based background regions produced with \texttt{ae\_better\_backgrounds}. This iterative (and rather computationally expensive) method attempts to achieve a balance between sufficient background counts and ``compactness'' of the region. Additionally, it considers the impact from the PSFs of nearby sources on a background region through modeling of the PSF. We also made sure that the backgrounds are always extracted from the same chip as the source.

We merged observations, extracted source products, and performed photometry in five different bands (0.5--8, 0.5--2, 2--8, 0.5--10, 1--10 keV) using \texttt{MERGE\_OBSERVATIONS} in AE, while choosing the flag \texttt{MERGE\_FOR\_PB}, which optimizes the merge for source validity. This option increases sensitivity to faint variable sources, at the cost of an increased detection significance for possibly false sources \citep{Broos10}.

Lastly, we used \texttt{FIT\_SPECTRA} to fit the spectra with an absorbed power-law. This fit was done as a quick benchmark and to obtain rough estimates on the absorbed flux in different bands. As described in \S\ref{sec:spec}, we perform robust spectral analysis through \textsc{BXA}.

\subsubsection{The special case of Terzan~5}
Terzan~5 stands out as a unique cluster in the X-rays. With the highest number of X-ray sources detected in any GC \citep[e.g.,][]{Heinke06b}, crowding poses a serious issue in the core of this cluster. Additionally, Terzan~5 contains the highest number of pulsars detected in any GC \citep[38 discovered so far; e.g.,][]{Lyne00,Ransom05,Cadelano18}. Several of these pulsars do have X-ray counterparts. Thus careful disentangling, detection, and localization of sources in this cluster is crucial. We note that \citet{Cheng19b} recently produced a careful study of the population and distribution of X-ray sources in Terzan~5 as this work was in the late stages of preparation. Given their focus on the radial distribution of sources, and our interest in careful consideration of MSP X-ray counterparts, we retain this cluster in our analysis.

For Terzan~5, we first corrected the absolute astrometry using the source catalog from \citet{Heinke06b}. For input source catalogs, we made four non-overlapping source catalogs. In addition to \texttt{wavdetect}, we also performed source detection using \textsc{PWDetect}\footnote{\url{http://cerere.astropa.unipa.it/progetti_ricerca/PWDetect/}} \citep{Damiani97}, as it is particularly powerful in the detection of faint sources in crowded regions. We then removed overlapping sources from \textsc{PWDetect} that had already been detected by \texttt{wavdetect}. We proceeded to compare this source catalog against the catalog of pulsars in Terzan~5 \citep{Ransom05}, and if a pulsar was not detected by \texttt{Wavdetect} or \texttt{PWDetect}, we added it to the input catalog for AE (a complete study of MSPs in Terzan~5 will be reported in Bogdanov et al.\ in prep). Lastly, we visually inspected the core of the cluster and added a total of 10 marginal sources, which were not picked up by the detection algorithms.

\subsection{Spectral analysis}\label{sec:spec}
After extracting/producing source and background spectra and associated files via AE, we performed a thorough and automated spectral analysis of all sources. First, we performed basic spectral fitting for all detected sources with an absorbed power-law using AE and Xspec \citep{Arnaud96} to naively estimate absorbed and unabsorbed flux for every source in multiple bands. However, for robust modeling, with careful consideration of uncertainties and the possibility of careful model comparison, we implemented the Bayesian framework provided by \textsc{BXA}\footnote{\url{https://johannesbuchner.github.io/BXA/}} \citep{Buchner14} which connects the nested sampling algorithm \textsc{MultiNest} \citep{Feroz19} with the fitting environment PyXspec\footnote{\url{https://heasarc.gsfc.nasa.gov/docs/xanadu/xspec/python/html/index.html}} \citep{Arnaud96} for spectral analysis. MultiNest allows efficient exploration of the parameter space in high dimensions which may contain multiple modes through uncorrelated samples.

We fit the stacked 0.3--10 keV X-ray spectrum of each source with W-stats statistics \citep{Cash79}. We binned each stacked spectrum to contain at least 1 background count per spectral bin, using the task \texttt{ftgrouppha} in \textsc{Heasoft}. This is to address a possible bias in some applications of W-stats \footnote{See Appendix B in the \textsc{Xspec} manual: \url{https://heasarc.gsfc.nasa.gov/xanadu/xspec/manual/XSappendixStatistics.html}}. We fit the spectra with three different models: absorbed power-law (TBABS$\times$PEGPWRLW), absorbed blackbody (TBABS$\times$CFLUX$\times$BBODYRAD), and absorbed diffuse gas emission (TBABS$\times$CFLUX$\times$APEC). Xspec's multiplicative component CFLUX was added for a robust constraint on unabsorbed flux (PEGPWRLW's normalization is already calibrated to yield model flux, thus CFLUX was not needed in that case). In all spectral analysis, we assumed \citet{Wilms00} abundances of elements and \citet{Verner96} photo-electric cross-sections. We assumed uniform priors on the power-law photon index (between $-1$ and 4) and the \texttt{lg10Flux} parameter in CFLUX (between $-17$ and $-8$). We assumed log-uniform prior for all other parameters to allow scale-invariance over multiple orders of magnitude: N$_\mathrm{H}$ between $10^{19}$ and $10^{24}$ cm$^{-2}$, $kT_{\text{APEC}}$ between 0.008 and 30 keV, and $kT_{\text{BB}}$ between 0.001 and 10 keV. Finally, we used model evidence ($\log_{10} Z$, as implemented in \textsc{BXA}) to compare the three different models for each source. 

\section{Catalog of X-ray Sources in Galactic Globular Clusters}\label{sec:catalog}
We produced a catalog of X-ray sources in 38 GCs through the steps explained in \S\ref{sec:data} and \S\ref{sec:analysis}, with measurements of numerous observational properties of these sources. The complete catalog is available in the electronic version of this article in Appendix \ref{sec:columns}. In Appendix \ref{sec:columns}, we also provide a complete overview of catalog entries/column available for each sources. In this section, we describe in detail how the catalog was developed and vetted, and provide an overview of various observed properties for cluster sources.

\subsection{Catalog Depth/Sensitivity}\label{sec:depth}
\begin{figure}
\centering
\includegraphics[scale=0.55]{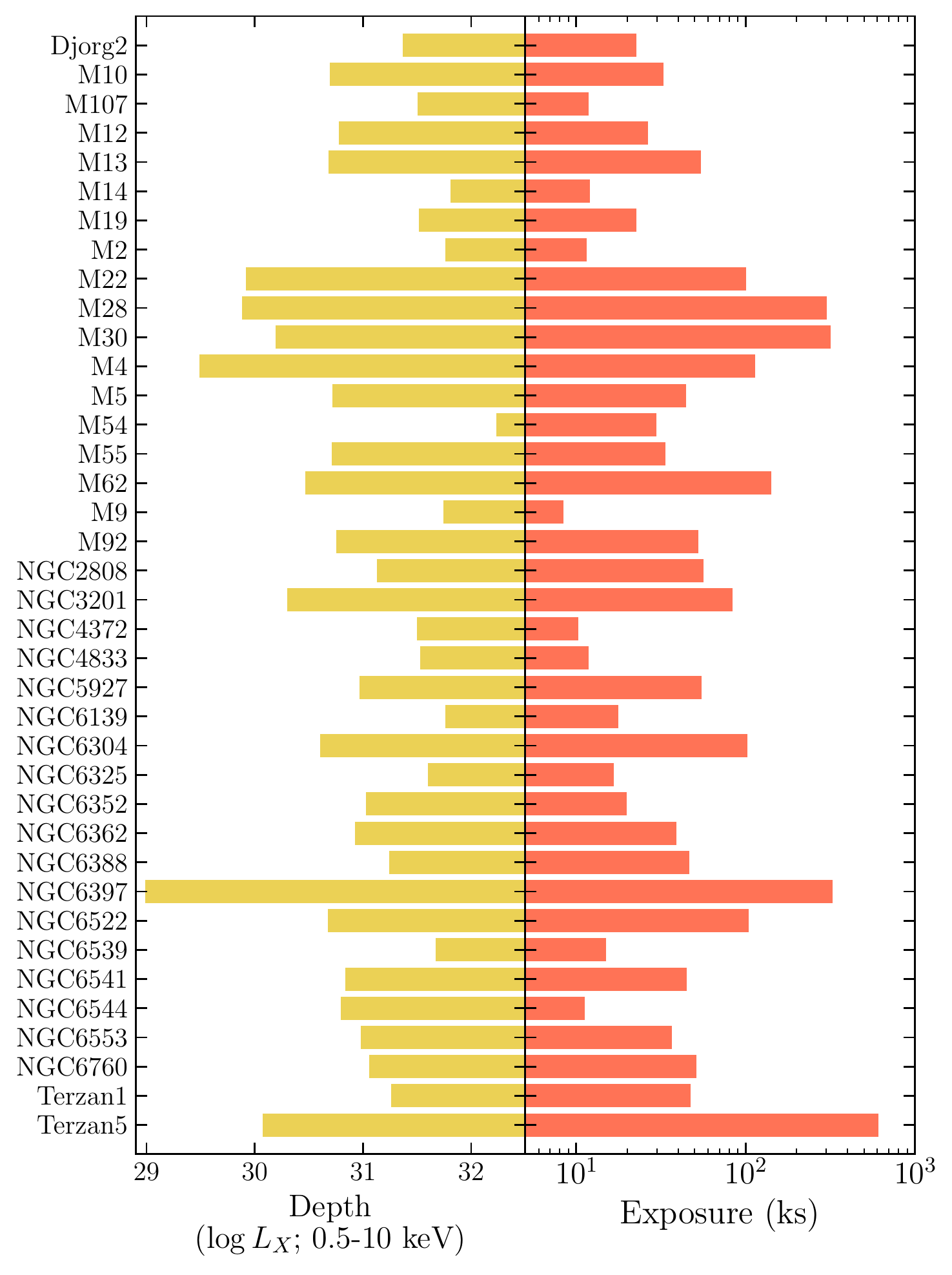}
\caption{Coverage and ``depth'' of \acis\ observations for the clusters in our sample. Depth is calculated based on X-ray luminosity (in \ergs) of a hypothetical source with 5 net counts, with an absorbed power-law spectrum (with the GC's N$_\mathrm{H}$ and a photon index of 1.7), over the total GC exposure time.}
\label{fig:depth}
\end{figure}

GCs in our sample are not observed uniformly in the X-rays. Some of these clusters have been observed for hundreds of kiloseconds, while some have only been observed for far shorter exposure time (Figure \ref{fig:depth}). This dispersion in depth only gets stronger when the impacts of variable interstellar absorption and distance are considered. Thus, it is important to estimate a sensitivity limit for each cluster taking the above factors into account.

We estimate a measure of X-ray flux sensitivity for each GC, assuming we can confidently detect an X-ray source with 5 net counts (in the 0.5--10 keV band) over the total exposure time for the GC with the primary CCD (\acis-S or \acis-I) in the primary \chandra\ cycle. We define primary CCD and cycle as the ones used in the majority of the observations. We need to consider the CCD and observing cycle that most of the GC observations were taken in because of sensitivity differences between \acis-S or \acis-I and impacts of contamination on \acis\ detectors over time\footnote{\url{https://cxc.harvard.edu/ciao/why/acisqecontamN0010.html}}. Thus, we use PIMMS\footnote{\url{https://asc.harvard.edu/toolkit/pimms.jsp}} to estimate the unabsorbed flux corresponding to a net count of 5 for each GC. We assume a typical power-law spectrum with a photon index of 1.7, and each cluster's expected hydrogen column density \citep[based on optical reddening values reported in the Harris catalog, and following the correlation reported by][]{Bahramian15,Foight16}. Finally, we use the GC distance to estimate the corresponding X-ray luminosity. These depth estimations are tabulated in Table~\ref{tab:sensitivity}, and are demonstrated in Figure \ref{fig:depth}.
The median depth is $L_X \sim 9\times10^{30}$ \ergs.

It is worth noting that these estimates are just first-order approximations, and are not hard cutoff thresholds. Variable backgrounds, off-axis angles, and spectral shapes are issues that this estimation does not address. However, these issues are unlikely to have large impacts in most cases. The GC most vulnerable to these issues is Terzan~5, where the high absorption severely extinguishes the sources at $L_X < 10^{31}$ \ergs\ \citep[and reduces the count rate for fainter sources more than brighter sources on average, since fainter sources tend to be softer, e.g.][]{Heinke06b}, and the crowding in the core increases the probability of missing sources in the high local background caused by the PSF wings of nearby sources.

\begin{deluxetable*}{lcccccccc}
\tablecaption{Estimation of depth and sensitivity for the GCs in our sample.}
\label{tab:sensitivity}
\tablehead{
\colhead{GC} & \colhead{Exposure} & \colhead{Prim. CCD} & \colhead{Prim. Cycle}  & \colhead{GC N$_\mathrm{H}$}   & \colhead{GC distance}& \colhead{Lim.~Unabs.~Flux}& \colhead{Lim.~L$_X$}  &  \colhead{\#Sources} \\
\colhead{}   & \colhead{(ks)}     & \colhead{}          & \colhead{}             & \colhead{(cm$^{-2}$)}& \colhead{(kpc)}               & \colhead{(\ergcms)}       & \colhead{(\ergs)}     &  \colhead{}
}
\startdata
Djorg~2  & 22.7     & 7           & 17  & $8.19\times10^{21}$ & 6.3   & $4.91\times10^{-15}$ & $2.32\times10^{31}$  & 2                \\
M~2      & 11.5     & 7           & 09  & $5.23\times10^{20}$ & 11.5  & $3.64\times10^{-15}$ & $5.75\times10^{31}$  & 5                \\
M~4      & 113.9    & 6,7         & 08  & $3.05\times10^{21}$ & 2.2   & $5.37\times10^{-16}$ & $3.10\times10^{29}$  & 100              \\
M~5      & 44.7     & 7           & 03  & $2.61\times10^{20}$ & 7.5   & $7.84\times10^{-16}$ & $5.26\times10^{30}$  & 13               \\
M~9      & 8.4      & 7           & 09  & $3.31\times10^{21}$ & 7.9   & $7.48\times10^{-15}$ & $5.56\times10^{31}$  & 1                \\
M~10     & 32.6     & 7           & 16  & $2.44\times10^{21}$ & 4.4   & $2.13\times10^{-15}$ & $4.91\times10^{30}$  & 13               \\
M~12     & 26.6     & 7           & 05  & $1.66\times10^{21}$ & 4.8   & $2.19\times10^{-15}$ & $6.02\times10^{30}$  & 6                \\
M~13     & 54.7     & 7           & 06  & $1.74\times10^{20}$ & 7.1   & $8.02\times10^{-16}$ & $4.82\times10^{30}$  & 18               \\
M~14     & 12.1     & 7           & 09  & $5.23\times10^{21}$ & 9.3   & $6.26\times10^{-15}$ & $6.45\times10^{31}$  & 5                \\
M~19     & 22.7     & 7           & 17  & $3.31\times10^{21}$ & 8.8   & $3.56\times10^{-15}$ & $3.29\times10^{31}$  & 4                \\
M~22     & 100.7    & 6,7         & 15  & $2.96\times10^{21}$ & 3.2   & $6.77\times10^{-16}$ & $8.26\times10^{29}$  & 80               \\
M~28     & 300.5    & 7           & 09  & $3.48\times10^{21}$ & 5.5   & $2.13\times10^{-16}$ & $7.66\times10^{29}$  & 136              \\
M~30     & 307.5    & 7           & 18  & $2.61\times10^{20}$ & 8.1   & $2.00\times10^{-16}$ & $1.57\times10^{30}$  & 20               \\
M~54     & 29.8     & 3           & 04  & $1.31\times10^{21}$ & 26.5  & $2.04\times10^{-15}$ & $1.70\times10^{32}$  & 7                \\
M~55     & 33.7     & 6,7         & 05  & $6.97\times10^{20}$ & 5.4   & $1.48\times10^{-15}$ & $5.15\times10^{30}$  & 17               \\
M~62     & 142.4    & 7           & 15  & $4.09\times10^{21}$ & 6.8   & $5.33\times10^{-16}$ & $2.94\times10^{30}$  & 74               \\
M~92     & 52.6     & 7           & 04  & $1.74\times10^{20}$ & 8.3   & $6.88\times10^{-16}$ & $5.65\times10^{30}$  & 12               \\
M~107    & 11.8     & 6,7         & 17  & $2.87\times10^{21}$ & 6.4   & $6.58\times10^{-15}$ & $3.21\times10^{31}$  & 4                \\
NGC~2808 & 56.2     & 1,3         & 08  & $1.92\times10^{21}$ & 9.6   & $1.23\times10^{-15}$ & $1.36\times10^{31}$  & 13               \\
NGC~3201 & 83.5     & 7           & 11  & $2.09\times10^{21}$ & 4.9   & $7.01\times10^{-16}$ & $2.00\times10^{30}$  & 32               \\
NGC~4372 & 10.3     & 6,7         & 17  & $3.40\times10^{21}$ & 5.8   & $7.91\times10^{-15}$ & $3.17\times10^{31}$  & 7                \\
NGC~4833 & 11.8     & 7           & 17  & $2.79\times10^{21}$ & 6.6   & $6.53\times10^{-15}$ & $3.39\times10^{31}$  & 2                \\
NGC~5927 & 54.5     & 7           & 13  & $3.92\times10^{21}$ & 7.7   & $1.31\times10^{-15}$ & $9.26\times10^{30}$  & 11               \\
NGC~6139 & 17.7     & 7           & 09  & $6.53\times10^{21}$ & 10.1  & $4.72\times10^{-15}$ & $5.74\times10^{31}$  & 7                \\
NGC~6304 & 102.5    & 1,2,3       & 11  & $4.70\times10^{21}$ & 5.9   & $9.65\times10^{-16}$ & $4.00\times10^{30}$  & 23               \\
NGC~6325 & 16.7     & 7           & 09  & $7.93\times10^{21}$ & 7.8   & $5.50\times10^{-15}$ & $3.99\times10^{31}$  & 5                \\
NGC~6352 & 19.8     & 7           & 13  & $1.92\times10^{21}$ & 5.6   & $2.86\times10^{-15}$ & $1.07\times10^{31}$  & 10               \\
NGC~6362 & 38.7     & 7           & 11  & $7.84\times10^{20}$ & 7.6   & $1.23\times10^{-15}$ & $8.49\times10^{30}$  & 17               \\
NGC~6388 & 46.7     & 7           & 06  & $3.22\times10^{21}$ & 9.9   & $1.49\times10^{-15}$ & $1.74\times10^{31}$  & 36               \\
NGC~6397 & 325.1    & 0,6,7       & 08  & $1.57\times10^{21}$ & 2.3   & $1.55\times10^{-16}$ & $9.76\times10^{28}$  & 124              \\
NGC~6522 & 104.0    & 2,7         & 04  & $4.18\times10^{21}$ & 7.7   & $6.74\times10^{-16}$ & $4.76\times10^{30}$  & 10               \\
NGC~6539 & 15.0     & 7           & 09  & $8.89\times10^{21}$ & 7.8   & $6.48\times10^{-15}$ & $4.70\times10^{31}$  & 7                \\
NGC~6541 & 44.8     & 7           & 04  & $1.22\times10^{21}$ & 7.5   & $1.02\times10^{-15}$ & $6.83\times10^{30}$  & 27               \\
NGC~6544 & 16.3     & 7           & 06  & $6.62\times10^{21}$ & 3.0   & $5.78\times10^{-15}$ & $6.20\times10^{30}$  & 6                \\
NGC~6553 & 36.7     & 7           & 13  & $5.49\times10^{21}$ & 6.0   & $2.23\times10^{-15}$ & $9.57\times10^{30}$  & 18               \\
NGC~6760 & 51.4     & 7           & 13  & $6.71\times10^{21}$ & 7.4   & $1.75\times10^{-15}$ & $1.14\times10^{31}$  & 12               \\
Terzan~1 & 47.3     & 6,7         & 17  & $1.73\times10^{22}$ & 6.7   & $3.41\times10^{-15}$ & $1.82\times10^{31}$  & 67               \\
Terzan~5 & 597.0    & 7           & 15  & $2.07\times10^{22}$ & 5.9   & $2.85\times10^{-16}$ & $1.18\times10^{30}$  & 188              \\
\enddata
\tablecomments{Primary CCD and primary cycle refer to \acis\ CCD (ACIS-I or ACIS-S) and \chandra\ observation cycle which most/all of a GC's observations were taken in. Cluster N$_\mathrm{H}$ and distance values are based on the Harris catalog. The limits on unabsorbed flux and X-ray luminosity are estimated in the 0.5--10 keV band for a source with 5 net counts over the GC total exposure, assuming a power-law spectrum with a photon index of 1.7 (see text). \#Sources represents the number of confident sources identified in each cluster (above the estimated threshold within the extraction region for each cluster, $\sim1.2\times R_{\textrm{hl}}$).}
\end{deluxetable*}

\subsection{Source Validity}\label{sec:validity}
As mentioned in \S\ref{sec:detection}, we chose moderately liberal detection thresholds in our initial source detection with \texttt{wavdetect} (with \texttt{sigthresh} typically set to 10$^{-4}$ or 10$^{-5}$), with a plan to better estimate our confidence
of source detection with AE, and carefully consider the PSF and local background in this second round. AE estimates source ``significance'' and a measure of false detection probability (``prob\_no\_source'') based on total source counts and background counts in the Poisson regime, following the method by \citet{Weisskopf07}. We estimated this probability in five different bands (0.5--8, 0.5--2, 2--8, 0.5--10, and 1--10 keV) for each source to increase sensitivity to faint sources that are extremely soft or hard. It is important to note that there are no simple ``number of independent trials'' that can be associated with the significance and probability measured this way by AE\footnote{AE manual, Section 5.10.3: \url{http://personal.psu.edu/psb6/TARA/ae_users_guide.html}}.

\begin{figure}
\includegraphics[scale=0.6]{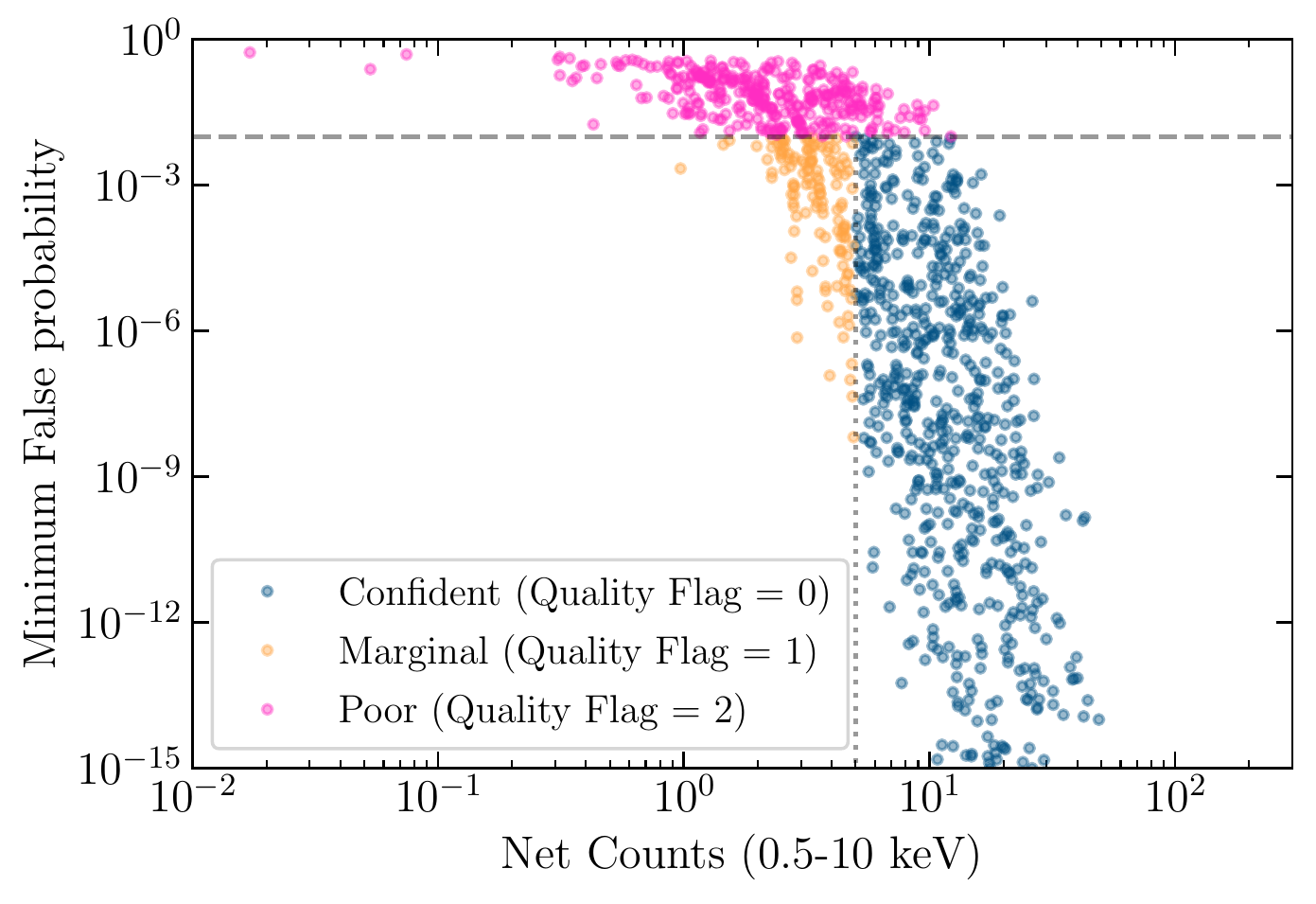}
\caption{Detection quality thresholds in the catalog. With these thresholds, our catalog contains 1139 confident, 134 marginal, and 394 poor detections. Note that there are numerous sources outside the plotted range with smaller ($\approx0$) false probabilities.}
\label{fig:detection}
\end{figure}

We do not remove any of the originally detected sources from our catalog. However, based on source net counts (in the 0.5--10 keV band) and the minimum false probability (minimum probability across the different bands), we assign a detection quality flag to each source. If a source has a minimum false probability value of $\geq 1\%$, we classify it as a poor detection (with a detection quality flag value of 2). If a source has a minimum false probability value of $< 1\%$ and a net source count $< 5$ (in the 0.5--10 keV band), we classify it as a marginal detection (detection quality flag value = 1). Finally, if a source has a minimum false probability value of $< 1\%$ and a net source count $\geq 5$, we classify it as a confident detection (detection quality flag value = 0; Figure~\ref{fig:detection}). The choice for a hard threshold of 5 on the net counts is partially arbitrary. However, it is somewhat informed by the fact that it is uncommon for a \chandra\ detection with $<5$ net counts to be robust without any other prior evidence (especially in the presence of strong background). Additionally, this choice allows estimates of reasonable flux sensitivity floors for each GC, assuming a minimum number of counts and given the accumulated \chandra\ exposure of the cluster (as detailed in \S\ref{sec:depth}).

Marginal and poor detections provide useful information about the sensitivity and robustness of our catalog. Furthermore, the false probabilities of these sources may be re-evaluated in the presence of external evidence (e.g., evidence of an interesting source at a given position in multi-wavelength observations).

With these classifications, this catalog contains 1139 confident, 134 marginal, and 394 poor detections. These classifications are reflected in the catalog under the column ``detection\_quality\_flag'', and source significance and false probablities in all five bands are reported under columns ``significance'' and ``prob\_no\_source'', respectively.

\subsection{Source Coverage and Extraction}\label{sec:locate}
For most of the sources in the catalog, all observations of their host GC cover these sources in a uniform fashion. However, some sources reside on the outskirts of large GCs, can fall on the chip gap (or fall off subarrays) in one or more observations, or can be impacted by large off-axis angles. Thus it is important to carefully consider which observations to merge to obtain optimized products. AE allows a systematic source-by-source merge of overlapping observations to allow optimization for a specific goal (e.g., maximizing source validity, enhancing source position precision, or enhancing the signal-to-noise ratio for photometry and spectroscopy) in the \texttt{MERGE} stage. We chose to merge aiming to maximize source validity. For most of the sources in the catalog (1429 out of 1667), this choice yields a stack of all observations covering an individual source. The total number of observations covering each source and number of observations used to produce merged products for each source are reported under columns ``num\_obs\_total'' and ``num\_obs\_merged''.

\begin{figure}
\includegraphics[scale=0.6]{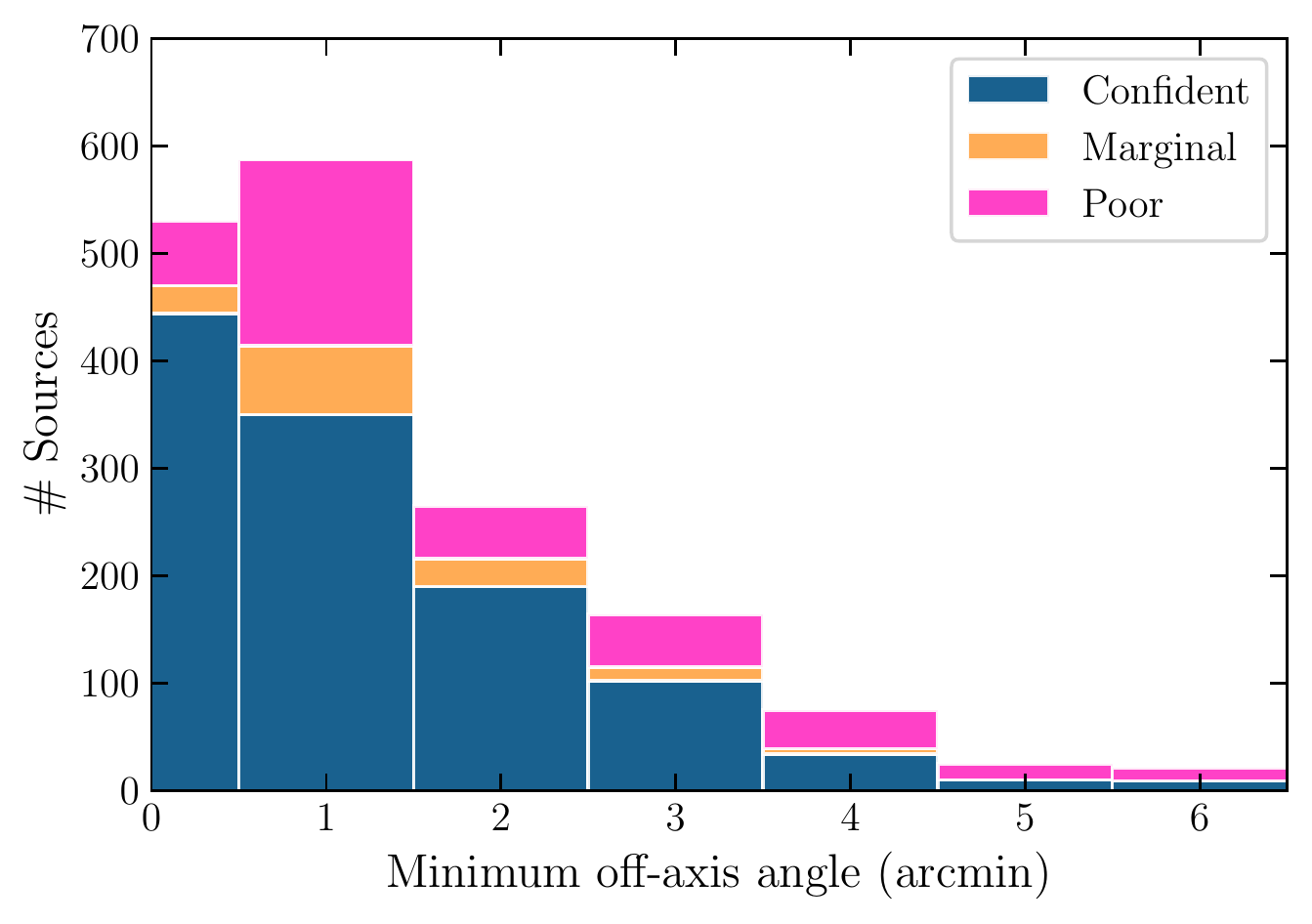}
\caption{Distribution of minimum off-axis angles in our catalog. 17\% of all sources in the catalog are observed only at off-axis angles larger than 2.5 arcmins.}
\label{fig:thetadist}
\end{figure}

We also include the range of off-axis angles at which each source has been observed (under columns ``theta\_low'', ``theta\_avg'', and ``theta\_high''), total on-source exposure time, and the primary \acis\ CCD covering each source (which is ACIS-S3 for all but 156 sources). Most of the sources in our catalog are observed at low off-axis angles at least once, however $\sim17\%$ of the sources have been observed only at off-axis angles larger than 2.5 arcmins\footnote{An off-axis angle of 2.5 arcmins is roughly where distortion of the \acis\ PSF becomes significant (e.g., see \url{https://cxc.harvard.edu/ciao/PSFs/psf_central.html}).} (Figure~\ref{fig:thetadist}). 

Off-axis angle and primary CCD are important attributes to consider when investigating individual sources. As we discuss in the next section, large off-axis angles can produce additional uncertainty in localization.

\subsection{Source Localization and Astrometry}
AE provides three methods to estimate source localization. The most basic method is via mean position of events in a specified band, with uncertainties estimated as the standard error of the mean. AE developers note that the mean position calculated this way\footnote{AE manual, section 5.3} may be biased when the PSF is asymmetric (e.g., at high off-axis angles) or when the background is non-uniform and ``sloped'' (e.g., when near a bright source). AE provides two other methods to estimate source position. One of these methods estimates the position by correlating the neighborhood around the source with source's PSF. This method may also be biased in the presence of non-uniform local background. Thus, AE provides a third method for estimating position based on image reconstruction and peaks. Unfortunately, currently there are no estimation of uncertainties available for correlation- and reconstruction- based localization methods. 

\begin{figure*}
\includegraphics[scale=0.75]{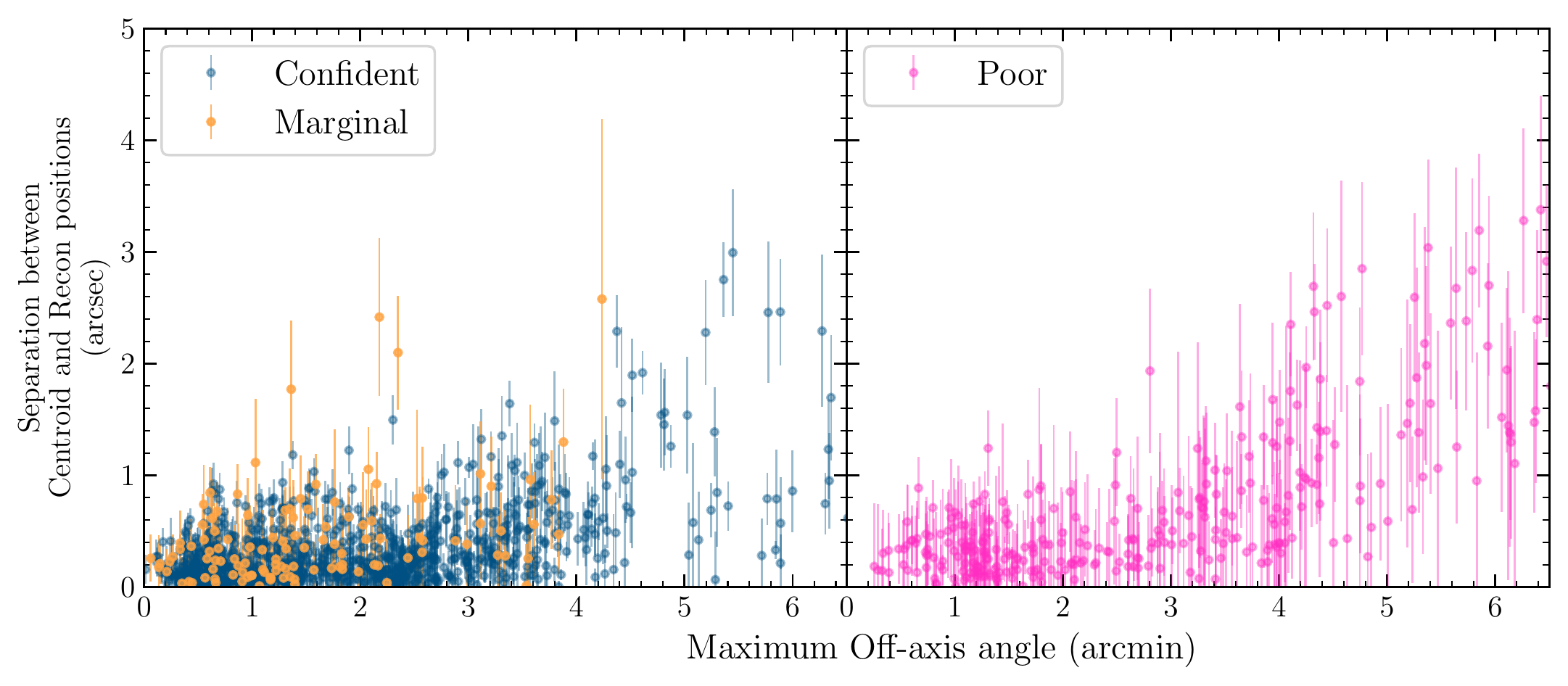}
\caption{Discrepancy between localizations based on centroiding and reconstruction versus off-axis angles. The discrepancy is negligible at low off-axis angles, but becomes important at large off-axis angles. The error bars plotted on the y-axis are the uncertainties on centroid localization.}
\label{fig:localization}
\end{figure*}

Given the large number of sources in the catalog and the spread in data quality, background, crowding, and off-axis angles, it is difficult to systematically choose a best localization method for each source. Thus we provide localization for each source via all three methods in the catalog. It is worth noting that while correlation- and reconstruction- based localizations are expected to provide more accurate estimates, especially at high off-axis angles, we notice that this may not be the case all the time. Therefore, we chose to extract products (light curve, spectra, etc.) for all sources based on the mean positions (the first method provided by AE). In most cases, the discrepancy between different localization methods has a negligible impact on these products (especially for confident detections), but not always. As demonstrated in Figure~\ref{fig:localization}, discrepancies between different localization methods are negligible at low off-axis angles for sources with sufficient counts. However, at large off-axis angles, this issue becomes important.

Thus, we urge the user to check localizations for their sources of interest, along with the used extraction region, source's off-axis angle, and crowding in the vicinity of the source. As described in \S\ref{sec:flags}, we provide figures and plots for each source to check these visually for each source in the catalog. 

\subsection{Photometry and Variability}\label{sec:phot/var}
We perform photometry (and source validation) in five bands (0.5--2,  2--8, 0.5--8, 0.5--10, 1--10 keV) for all sources using AE. As discussed in detail in the AE documentation (section 5.6.3), adjusting the background scaling is a crucial step to gain photometric accuracy and more robust source validity. We follow the suggestions for this technique and re-adjust the background scale iteratively, and then perform photometry and source validation. We report source counts, background counts, background scaling factor, background-subtracted ``net'' counts, and uncertainties on the net counts as estimated following \citet{Weisskopf07} in all the aforementioned bands. 

AE also provides Kolmogorov-Smirnov \citep[KS;][]{Kolmogorov33,Smirnov48} and $\chi^2$ tests, and their associated p-values, for source variability, within and between observations. Given that a majority of our sources are in the low-count regime (and thus do not have Gaussian uncertainties on count rates), we neglect the $\chi^2$ test altogether, and only include the KS test p-values for variability within and between observations in the catalog. The p-values for variability within an observation are the minimum value among the observations. P-values closer to 0 indicate higher probability of variability. It is worth noting that large P-values merely indicate inconclusive evidence on variability in the examined data. As we warn in \S\ref{sec:caveats}, these values should be interpreted with caution. 

\subsection{Spectral Analysis}
We provide the results of our spectral analysis (as explained in \S\ref{sec:spec}). These results include absorbed and unabsorbed flux values in different bands as estimated via AE (through the \texttt{flux} task in Xspec), assuming a simple power-law fit. We also list the median and 10\% and 90\% quantiles of the parameter posterior distributions resulting from fitting three separate models: an absorbed power-law, absorbed blackbody, and absorbed diffuse plasma (APEC) in \textsc{BXA} and \textsc{Xspec}. We also provide unabsorbed X-ray flux and X-ray luminosity (assuming the source is in the GC) for each model in the 0.5--10 keV band. Finally, we give the relative model probability (with the probability for the most likely model set to 1.0) as estimated based on model evidence, and indicate the best model based on these probabilities.

For completeness and comparison, we also provide the expected hydrogen column density towards the host GC, estimated based on the correlation between N$_\mathrm{H}$ and $A_V$ \citep{Bahramian15, Foight16}, using the $E(B-V)$ values from the Harris catalog and assuming $R_V = 3.1$. We also estimate the uncertainties on the GC N$_\mathrm{H}$ values, considering the reported uncertainties on $E(B-V)$ in the Harris catalog, the reported uncertainty on the correlation slope by \citet{Bahramian15}, and assuming an uncertainty of 0.1 on $R_V$. We do not use these values for fitting, but as discussed in the next section, they provide a naive indicator to distinguish some foreground sources.

\begin{figure*}
\includegraphics[scale=0.45]{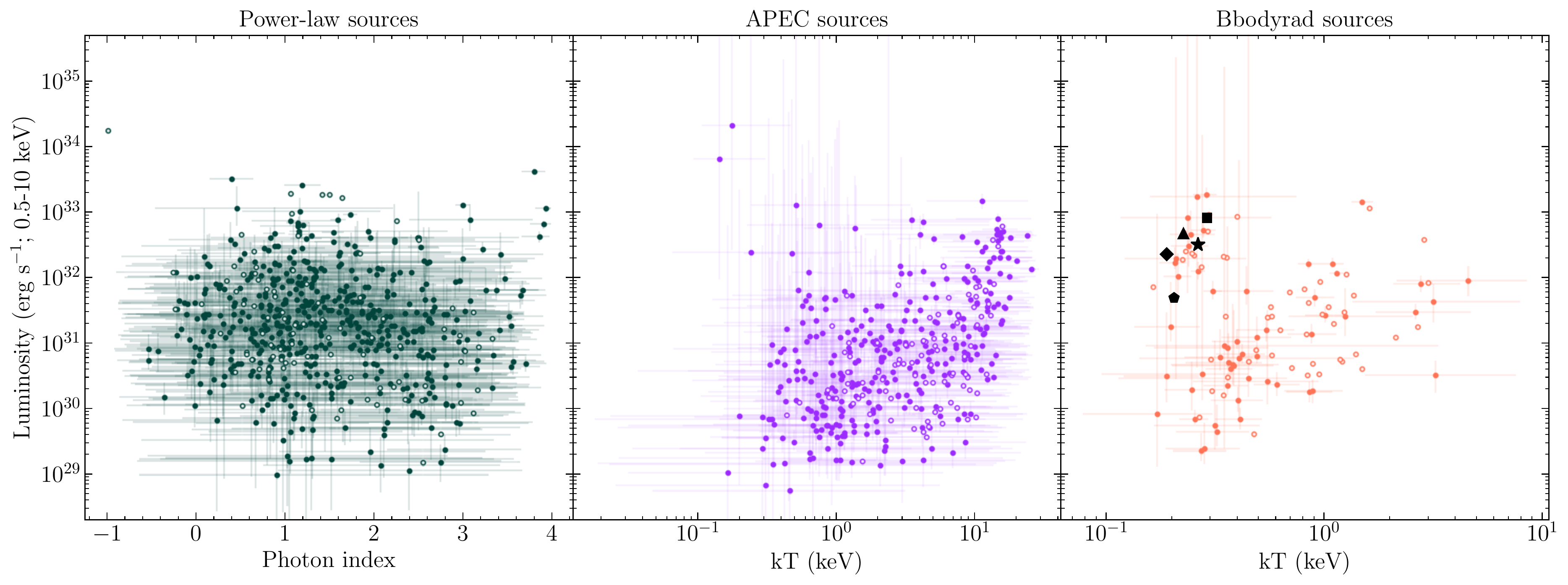}
\caption{Overview of the spectral fitting results for the confident sources. Each source is plotted only in the panel representing the best-fit model. Luminosities are estimated assuming the source is at its host GC distance. Hollow points represent possible foreground sources, based on comparing the best-fit N$_\mathrm{H}$ for the source to the N$_\mathrm{H}$ expected for the GC (see \S\ref{sec:flags}). There are 620 sources best fit with a power-law, 409 sources best fit by an APEC model, and 110 sources best fit by a blackbody. The black markers represent some of the known quiescent NS-LMXBs in our sample, including CXOU J214022.17--231046.2 in M~30 (black triangle), CXOU J164143.76+362757.9 in M~13 (black diamond), CXOU J182432.81--245208.5 in M~28 (black square), CXOU J174041.48--534004.5 in NGC~6397 (black pentagon), and CXOU J171432.95--292748.1 in NGC~6304 (black star).}
\label{fig:goose_spectra}
\end{figure*}

The results of spectral fitting for the confident detections are plotted in Figure~\ref{fig:goose_spectra}. It is worth noting that the sources with blackbody as the best-fit model include some of the well-studied known quiescent NS-LMXBs \citep[e.g., see][]{Guillot13, Steiner18} among dozens of other sources, indicating a great potential in this model selection for identification and study of new NS-LMXBs.

\subsection{Flags, GC-associated aspects, and convenience plots}\label{sec:flags}
We provide a number of columns in the final catalog to allow easy assessment of source conditions. The column ``detection\_quality\_flag'' indicates the robustness of the detection based on source net counts and validity (as explained in \S\ref{sec:validity}). The ``spectrum quality flag'' is merely based on total number of source counts: if a source has $\gsim100$ counts, it would have relatively reliable spectral analysis (and we assign a flag value of 0). If the total number of counts is somewhere between $\lsim100$ and $\gsim20$, the estimates are less reliable and should be taken with caution (flag value of 1). Lastly, if a source has $\lsim20$ counts, spectral analysis is merely suggestive, and model comparison is not to be taken with confidence (flag value of 2).

We also provide a simple assessment of whether a source {\it could} be a foreground object. This assessment relies on comparison of the 90\% upper limit on a source's N$_\mathrm{H}$ (from the best-fit model) when compared to the GC's expected N$_\mathrm{H}$. If a source's N$_\mathrm{H}$ upper limit is lower than the GC N$_\mathrm{H}$, there is a chance that it may be a possible foreground object. We emphasize that this is only a zeroth-order suggestive indicator and a ``False'' value for the ``foreground\_flag'' does not rule out a foreground nature entirely, and a ``True'' value, especially for significantly bright sources, may be a result of incomplete/inaccurate spectral modeling. This flag is effective for identification of faint foreground objects in the direction of highly-absorbed GCs (like Terzan~5, e.g., with $\sim50$ possibly foreground sources), while it would not be a good indicator for low absorption ones (like M~30). 

Precise astrometry corrections are a challenge in most X-ray observations. At this point, most of the GCs in our catalog do not have absolute astrometry corrections applied, meaning that the user should consider an additional $\sim0\farcs8$ uncertainty based on the accuracy of \chandra's absolute astrometry\footnote{\url{https://cxc.harvard.edu/cal/ASPECT/celmon/}}. Currently, we have corrected absolute astrometry only for Terzan~5, following crossmatching sources to \citet{Heinke06b}. We hope to provide this correction for more clusters in future updates to the catalog. The status of astrometry correction is indicated for each source (each GC) under ``abs\_astrometry\_flag''.

\begin{figure}
\centering
\includegraphics[scale=0.6]{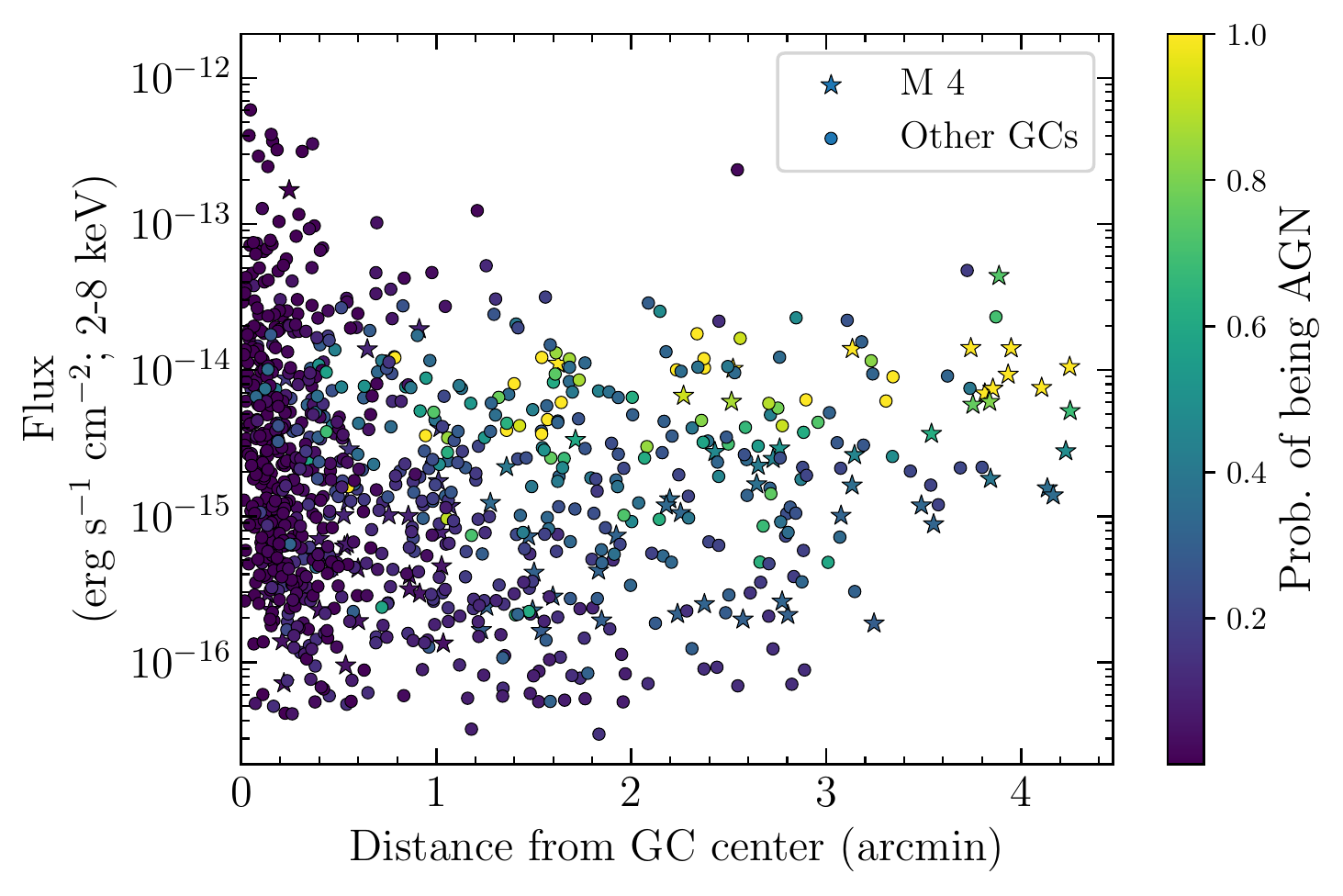}
\caption{For all sources in our catalog, absorbed flux is plotted  as a function of distance from the GC center, and points color-coded to denote the probability of being a background AGN. Sources in M~4 are highlighted, as due to its nearby distance (2.2 kpc) and large half-light radius ($4\farcm3$), a large number of background AGNs are expected in this cluster.}
\label{fig:agn_prob}
\end{figure}

Beside providing the host GC's name, distance, expected N$_\mathrm{H}$ and half-light radius (all based on the Harris catalog), we also provide each source's angular distance from the cluster center, along with the number of background Active Galactic Nuclei (AGNs) with higher flux than the source expected within the cluster half-light radius, using the approximation provided by \citet{Mateos08}. It is worth noting that study of background AGNs is mostly done in low-extinction parts of the sky, while some of the GCs in our sample have very high extinction values, making it difficult or impossible to detect the ``typical'' contaminating AGNs in these GCs. To reduce the impact of this bias, we use unabsorbed flux values for our sources when calculating the \citet{Mateos08} factor. The number of expected AGNs, along with GC half-light radius and source distance from GC center, provides a proxy for the likelihood that a source is a background AGN. We estimate a crude probability that each source is associated with the GC background, based on where the source is located in the GC (in the core or outside the core), number of expected AGNs with similar or higher flux in the 2--10 keV band ($N_{\rm{BKG}}(F_X>F_{X,\rm{src}})$), and the number of detected sources in that region with similar or higher flux ($N_{\rm{Detect}}(F_X>F_{X,\rm{src}})$). We defined the probability of being a background AGN as the ratio of $N_{\rm{BKG}}(F_X>F_{X,\rm{src}})/N_{\rm{Detect}}(F_X>F_{X,\rm{src}})$ for each source. If the source is located in the cluster core, $N_{\rm{BKG}}(F_X>F_{X,\rm{src}})$ and $N_{\rm{Detect}}(F_X>F_{X,\rm{src}})$ were estimated for the cluster core. If the source is outside the core but within the half-light radius, the estimate only includes the half-light region (excluding the core). This diagnostic impacts clusters with various apparent sizes differently. For example, the number of background AGNs detected in M~4, a nearby GC with a half-light radius of $4\farcm3$, which has  been observed for more than 100 ks with \acis, would be significantly higher than in most other GCs (Figure~\ref{fig:agn_prob}). It is also worth noting that for some clusters closest to the Galactic Plane, foreground sources may be a bigger contaminant than background sources.

\begin{figure*}
\centering
\includegraphics[width=0.86\textwidth]{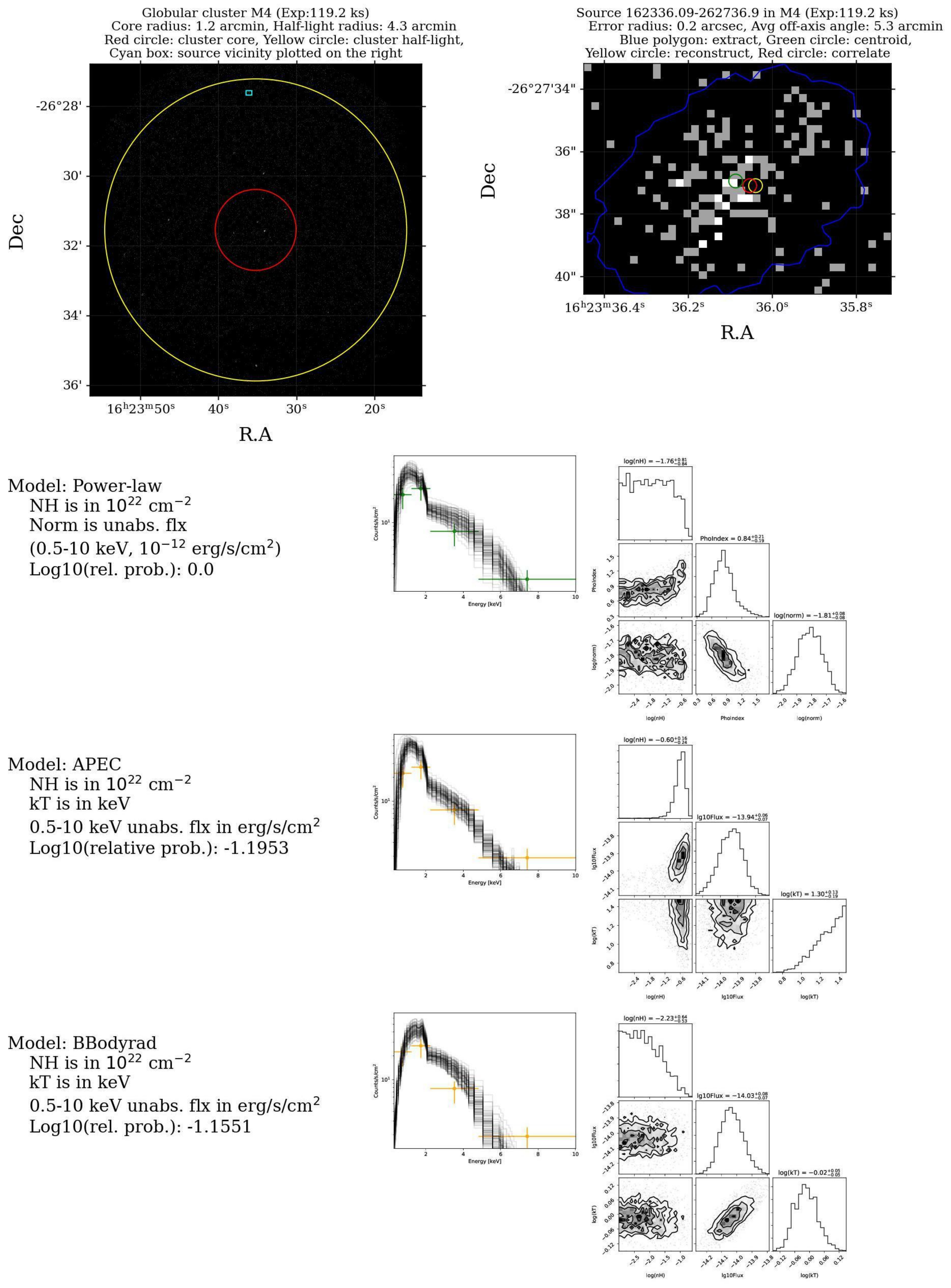}
\caption{An example of convenience plots for a catalog source at high off-axis angle in the cluster M~4. The top left panel shows the location of the source within the cluster, to allow assessment of membership and impact of artifacts (the red circle denotes the extent of the cluster core, the yellow circle display the cluster half-light radius, and the cyan box denotes the source region. These are superimposed on the \acis\ image). The top right panel is a zoom-in on the source, with various localization estimates marked (as green, yellow, and red circles for the centroiding, correlation and reconstruction techniques, respectively) and the extraction region pointed out as a blue polygon. The lower panels show results of spectral fitting for each of the three models (from top to bottom: power-law, blackbody, APEC). These include a representation of the binned spectrum and the posterior sample of fits, plus a corner plot showing the posterior distribution of model parameters. A complete and high-resolution figure-set including plots for all sources in the catalog is available in the online version of this article.}
\label{fig:convplot}
\end{figure*}

As an auxiliary resource, for each source we provide a compilation of plots summarizing its position within the GC and localization and extraction regions, along with plots showing the three spectral fits and parameter posterior distributions. These plots are meant to facilitate manual assessment of individual sources and are available as a complete figure-set in the online version of this article (see Figure~\ref{fig:convplot} for an example) and are also accessible at \url{https://bersavosh.github.io/research/goose.html}.

\section{Discussion}\label{sec:disc}
\subsection{Caveats, warnings and pitfalls}\label{sec:caveats}
The catalog presented in this paper is a result of multi-level semi-automated analysis of a large set of heterogeneous X-ray observations. Thus, there are shortcomings and caveats in the results, which we note below.

As discussed in \S\ref{sec:locate}, we provide multiple localizations for each source in the catalog based on available methods. It was infeasible in this work to determine the best method for each individual source without careful consideration of numerous factors (like position on the chip, crowding around the source, background profile, source brightness, off-axis angle). Thus, we extracted all high-level {\it source} products (spectra, light curves, and associated files), using the centroid position. As shown in Figure~\ref{fig:localization}, the difference between localization based on centroiding and image reconstruction is generally small enough that the impact on extraction of high-level products is negligible. However, users are cautioned to check the extraction regions (as presented in convenience plots like Figure~\ref{fig:convplot}), especially for sources at high off-axis angles.

The catalog contains some information regarding variability properties of individual sources. However, it is important to note that AE does not provide background-subtracted light curves and thus does not distinguish between source and background events. This issue strongly impacts light curves (and subsequently results of any variability tests) of sources with low numbers of counts, especially in crowded regions (e.g., the core of Terzan~5). Furthermore, while the KS-test is a strong distribution-free and easy-to-compute test, it has well-documented shortcomings \citep[e.g., poor sensitivity in the tails of distributions;][]{Babu04, Babu06, Feigelson12}. These issues become more significant in the photon-starved regime. Thus, we advise caution in interpreting the results of variability tests, and urge users to consider source brightness and possible impact of background and/or nearby sources on these results.

As discussed in \S\ref{sec:spec}, we used Bayesian nested sampling Monte-Carlo methods for spectral fitting through the \textsc{bxa} package on top of \textsc{Pyxspec}. Given the nature of this method, it is important to consider the prior constraints on a parameter (as detailed in \S\ref{sec:spec}) and the shape of the posterior distribution for interpretation (i.e., whether a parameter has hit the upper/lower limit implied by the prior). The parameter posteriors are not always unimodal or quasi-gaussian, and some parameters in a model can be strongly correlated. The auxiliary plots (available as an electronic figureset accompanying this article, also available at \url{https://bersavosh.github.io/research/goose.html}) include 1-D and 2-D histograms of posterior distributions for each model fit, to allow inspection of these issues.

It is also important to note that while our spectral fitting method provides a rather robust evaluation of parameter space, it will not overcome the intrinsic uncertainties in the extremely low-count statistics regime. For example, in some cases, spectral fitting at low S/N indicates an excessively high N$_\mathrm{H}$, coupled with a very high X-ray luminosity (while the absorbed flux is rather low). Thus, it is important to check the uncertainty bounds and the shape of the posterior distribution at low S/N, and we urge caution in interpretation of spectral modeling results and model comparison for sources with $< 20$ counts.

Lastly it is worth noting that the models used for spectral analysis here are simple single-component models and they may not be sufficiently descriptive for some sources. This is especially the case for the brightest sources with hundreds to thousands of counts; the high S/N spectra for such sources may demand multi-component models. Additional components may be necessary to represent system emission processes (e.g., a blackbody or a neutron star atmosphere model + power-law for a weakly accreting NS-LMXB, see \S\ref{sec:brights}). Also note that the photometric and spectral analysis presented in this catalog does not include pile up corrections. Thus, it is important to consider the potential impact of pile up when using results for bright sources. AE provides an assessment of whether a source may be piled up based on PSF simulations, and we have included a flag column (named ``pile up flag'') in the catalog indicating potential pile up. If this flag is ``True'' for a source (this occurs for only 15 sources), spectral analysis (and assertions/flags relying on that) presented in the catalog may be inaccurate.

\subsection{Population and X-ray luminosity function of faint X-ray sources in Galactic GCs}\label{sec:xlf}

The number of observed XRBs in a GC depends on a variety of both intrinsic and observational factors, ranging from the GC's XRB production rate to the depth of the observations and distance to the cluster. It has been established that the number of XRBs in a GC correlates strongly with the cluster encounter rate \citep{Pooley03, Bahramian13}. 

\begin{figure*}
\includegraphics[scale=0.7]{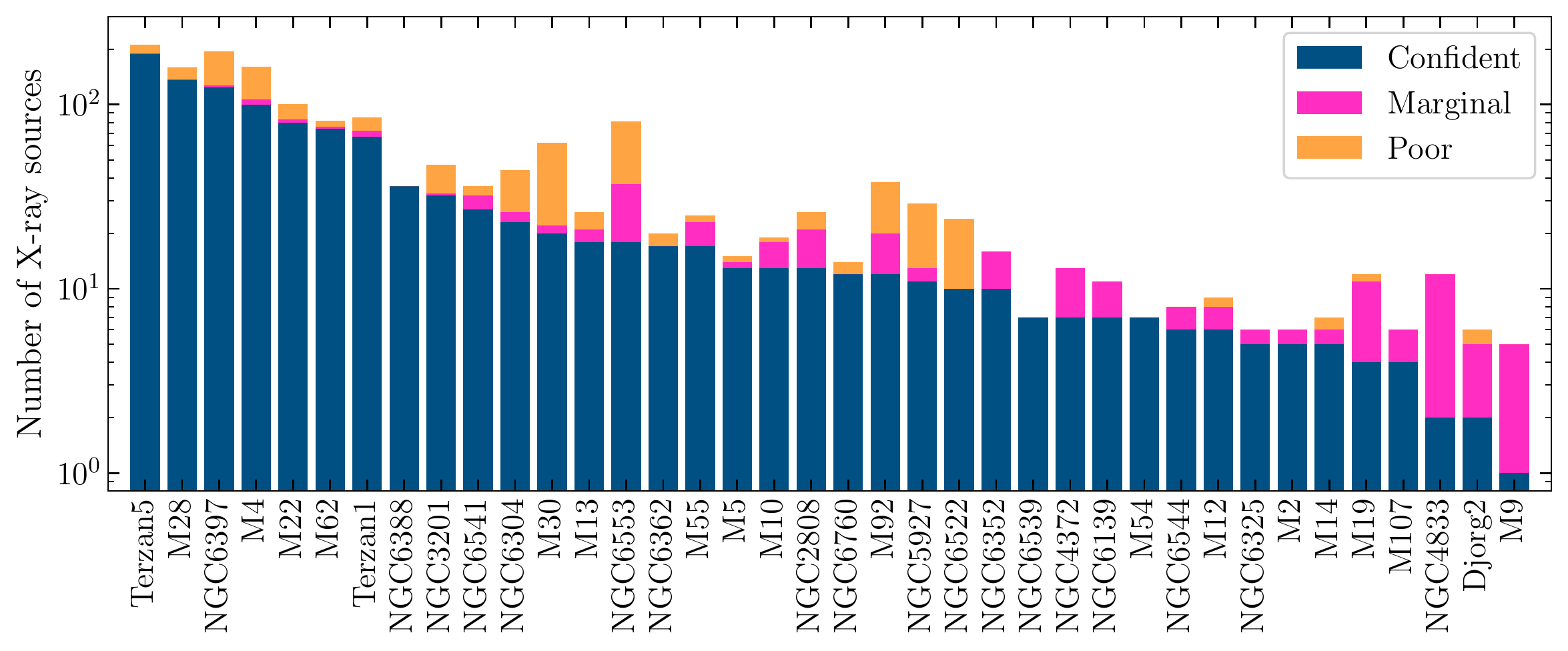}
\caption{The number of detected X-ray sources in our catalog as a function of cluster. The numbers include sources that are likely foreground/background objects (strongly impacting GCs like M~4). }
\label{fig:gc_hist}
\end{figure*}

We detect the largest number of X-ray sources towards Terzan~5, which has long been known to host a large number of XRBs, and has been observed extensively. Terzan~5 is followed by M~28 and NGC~6397, both of which show more than a hundred X-ray sources (Figure~\ref{fig:gc_hist}). 

\begin{figure}
\centering
\includegraphics[scale=0.6]{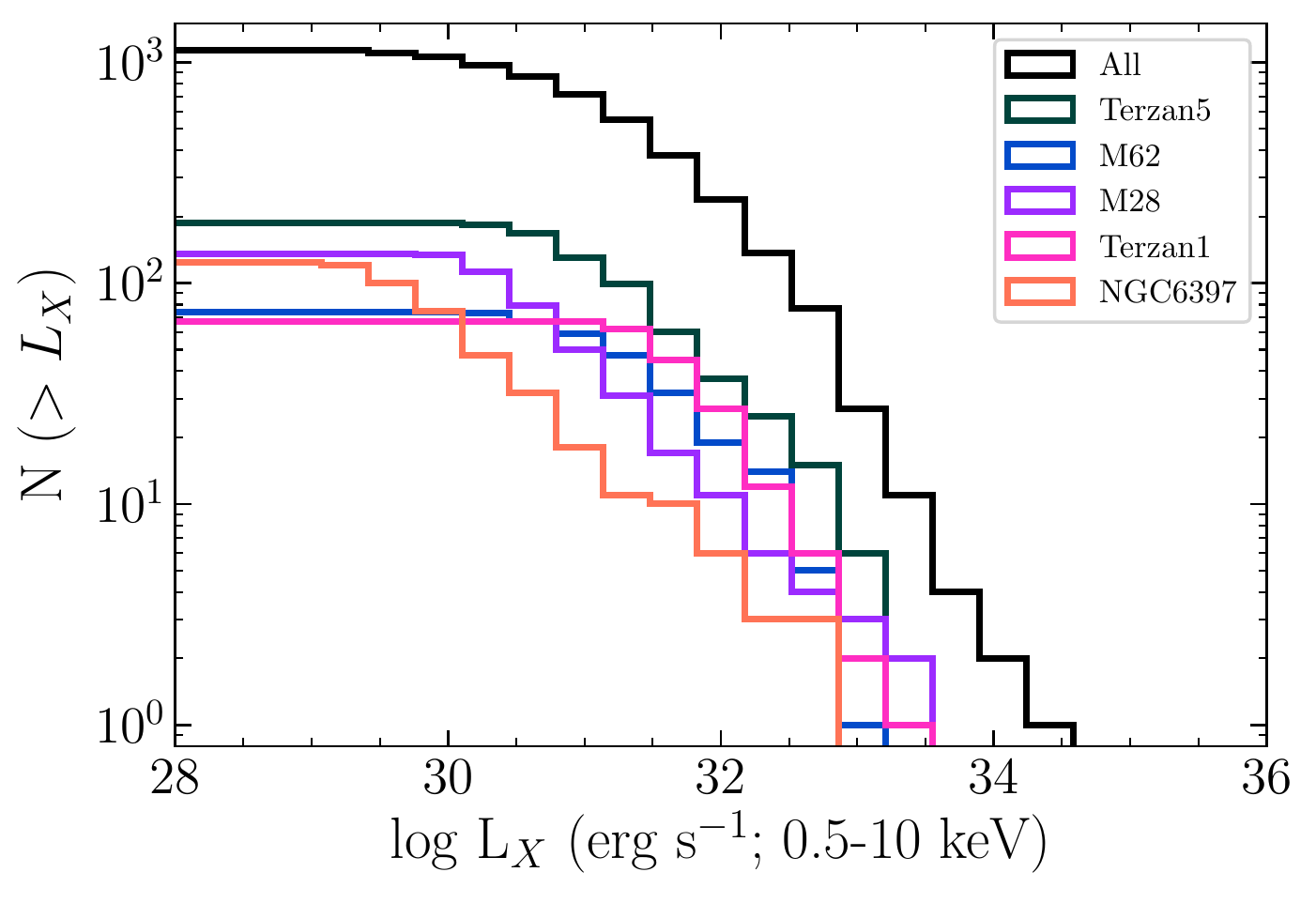}
\caption{The cumulative X-ray luminosity distribution of X-ray sources in GCs in this study is plotted as a solid black line.  Contributions of some noteworthy GCs are plottted as colored lines.}
\label{fig:xlf}
\end{figure}

Using our catalog, we estimate an empirical X-ray luminosity distribution for X-ray sources with $L_X < 10^{36}$ \ergs\  (Figure~\ref{fig:xlf}). For this purpose, we only consider confident detections (1139 sources). Since all GCs in our sample except for M~54 are nearly complete for $L_X > 10^{32}$ \ergs, the luminosity distribution can be considered close to complete down to that value, with substantial incompleteness present for $L_X \lesssim 10^{31}$ \ergs. The caveats to the completeness for $L_X > 10^{32}$ \ergs\ would be any sources with heavy absorption (including partial covering) not appropriately modeled with existing data, and the possibility of a small number of sources blended in the crowded cores of M~62 or Terzan~5.

However, in interpreting the luminosity distribution, it is important to consider 
observational biases such as the biased contribution of clusters with especially deep data. For example, sources with luminosities $< 10^{30}$ \ergs\ are not detected in Terzan~5 due to high extinction, even though the cluster has been observed extensively. By contrast, there are numerous low-luminosity sources in the sample from nearby, low-$N_H$ GCs like NGC~6397. It is also important to note that the luminosity distribution here is heavily impacted by foreground and background sources in the direction of each GC. While the foreground flag and AGN probability provided in this catalog could in principle help with pruning out foreground and background sources, they are both very rough assessments (see \S\ref{sec:flags}). 

\subsection{Nature of the brightest sources in the catalog}\label{sec:brights}
Typically, quiescent XRBs with main sequence companions are observed with X-ray luminosities $\lsim10^{33}$ \ergs. However, some XRBs show brighter ``quiescent'' luminosities ($10^{33} \lsim L_X \lsim 10^{36} $ \ergs). Accretion at these luminosities is less well understood.
Systems at these luminosities are sometimes called very faint X-ray binaries \citep[ e.g.,][]{Wijnands06,Degenaar09}. Such sources are likely to be a heterogeneous class of objects, including transitional millisecond pulsars in their subluminous disk states, symbiotic X-ray binaries (with giant companions), and other binaries with accretion at an unusually low or inefficient level for unknown reasons \citep[e.g.,][]{Bahramian19,Shaw20}.
An advantage of the current catalog in studying these XRBs is the well-constrained distances, compared to  field sources where the distance is ususally unknown. 

In this section we explore the nature of sources from our catalog that have confidently-measured X-ray luminosities $\geq10^{33}$ \ergs\ (Table~\ref{tab:bright}). It is worth reiterating that the luminosities estimated in our catalog are based on the stacked spectrum for each source and thus represent average luminosities based on \chandra\ observations; sources that have shown a brief excursion to high luminosity in a large set of observations are unlikely to be included. First, in section \S\ref{sec:brightknown} we will briefly summarize the $L_X > 10^{33}$ \ergs\ sources in our catalog that have been previously studied in detail, and in \S\ref{sec:brightnew}, we will discuss sources with $L_X > 10^{33}$ \ergs\  identified in our survey for the first time.

\subsubsection{Known Systems}\label{sec:brightknown}

CXOU J173617.42--444405.9 (IGR J17361--4441) is an X-ray transient source in NGC~6388, discovered by \integral\ in 2011 \citep{Gibaud11, Wijnands11}. It had an unusually low outburst peak $L_X < 10^{36}$ \ergs. Comparison of \chandra\ observations of NGC~6388 from before the outburst to the one taken during this source's outburst indicated a very faint quiescent level, with $L_X < 10^{31}$ \ergs\ \citep{Pooley11}. Based on X-ray spectroscopy and radio non-detections during the outburst, \citet{Bozzo11} concluded that this system is likely an NS-LMXB. There have been no further outbursts detected from this cluster to date.

CXOU J173545.56--302900.0 (Terzan~1 CX1) is a bright ($L_X\sim2\times10^{33}$ \ergs) X-ray source identified by \citet{Wijnands02a}. \citet{Cackett06a} showed that the source has a very hard X-ray spectrum and may be an intermediate polar with partial covering, i.e., an accreting magnetic white dwarf with a truncated accretion disk where a subset of the X-ray emission is absorbed.

CXOU J180150.32--274923.6 (OGLE-UCXB-01) is the brightest source in Djorg 2.  It has $L_X \sim 2\times 10^{33}$ \ergs\ and a hard X-ray spectrum. \citet{Pietrukowicz19} identified this X-ray source as the counterpart to a variable OGLE source with a periodic modulation of 12.8 minutes. Based on optical and X-ray properties of the system, they suggest the system might be an ultra-compact X-ray binary, harboring a NS (or a BH) accreting from a WD. However, an alternative scenario such as an intermediate polar is not ruled out.

CXOU J182432.00--245210.9 (PSR J1824--2452A) is one of seven radio MSPs in the core of M~28 that show X-ray emission, and is the X-ray brightest of these systems \citep{Bogdanov10b}. It is rather unusual for X-ray emission from radio pulsars to reach $L_X > 10^{33}$ \ergs: this was actually the first MSP discovered in a GC \citep{Lyne87}, and is among the most energetic MSPs known. \citet{Bogdanov10b} argue that the X-ray emission from this pulsar is likely due to heated magnetic polar caps.

CXOU J174805.23--244647.3 (EXO 1745--248) is a known transient X-ray burster in the core of Terzan~5, with an unusually high quiescent X-ray luminosity \citep{Wijnands05, Galloway08}. Study of its quiescent behavior over the last two decades with \chandra\ has shown extreme variability in its quiescent luminosity, ranging between $3\times10^{31}$ and $2\times10^{34}$ \ergs\ \citep{RiveraSandoval18a}.

CXOU J182432.49--245208.1 (IGR J18245--2452; PSR J1824--2452I) in M~28, is one of a handful of transitional MSP systems known (and the only one that has shown a bright X-ray outburst). The system was identified as a transitional MSP following a bright outburst in 2013, during which consistent pulsations were first detected in the X-rays, and then in the radio \citep{Papitto13}. So far, there have been three luminosity ``states'' observed in transitional MSPs, including a very faint ($L_X \leq 10^{32}$ \ergs) pulsar state, a variable subluminous ($L_X \sim 10^{33}-10^{34}$ \ergs) accretion state, and a bright ($L_X > 10^{36}$ \ergs) outburst state \citep[which has only been seen to date in IGR J18245--2452,][]{Linares14b}. \chandra\ observations of IGR J18245--2452 (which exclude the outburst) show strong variability between $10^{32}$ and $10^{34}$ \ergs; most of the time, the system is in the faint accretion state with $L_X > 10^{33}$ \ergs\ \citep{Linares14a}.

CXOU J174805.41--244637.6 is the third transient XRB discovered in the core of Terzan~5 \citep{Bahramian14}. Detection of X-ray bursts proved the NS-LMXB nature of this system \citep{Altamirano12atel}. Identification of the quiescent counterpart in \chandra\ observations allowed study of its long-term behavior pre- and post- outburst \citep{Homan12atel}, indicating that the system shows a persistently high X-ray luminosity ($> 10^{33}$ \ergs, both before and after the outburst; \citealt{Degenaar15}). This suggests that continuous low-level accretion is present in the system.

\subsubsection{New Systems}\label{sec:brightnew}
A handful of bright systems in Table~\ref{tab:bright} have not been previously studied in detail or in some cases were not previously known. For these sources, their fluxes are generally high enough to allow sufficient counts in their spectra for more complex modeling. Thus, in this section we investigate their X-ray behavior in more detail, and attempt to determine their nature with some help from radio data provided by the MAVERIC survey \citep{Shishkovsky18,Tremou18}, and optical catalogs like the ACS treasury \citep{Sarajedini07}. 

\begin{figure*}
\centering
\includegraphics[scale=0.45]{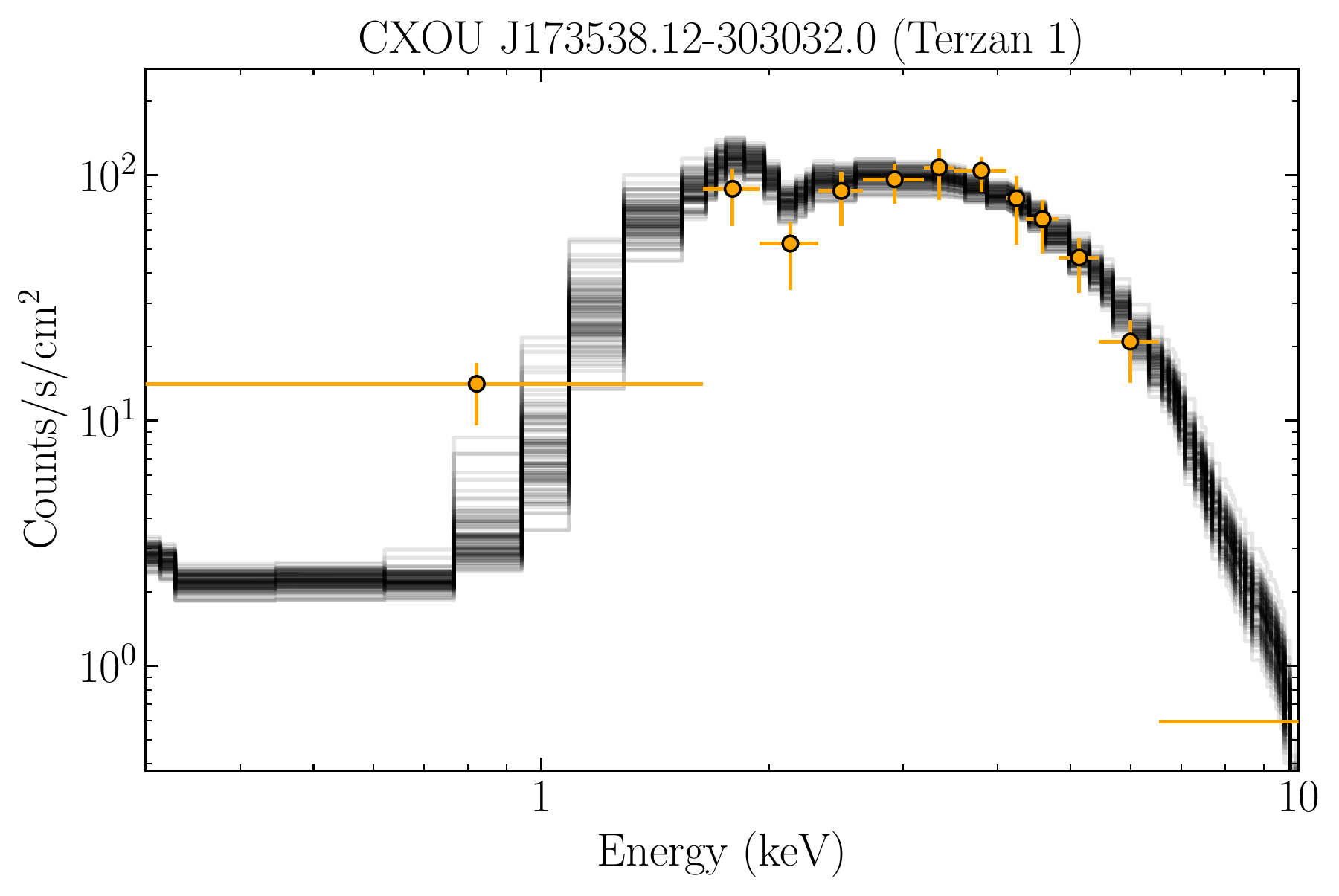}
\includegraphics[scale=0.45]{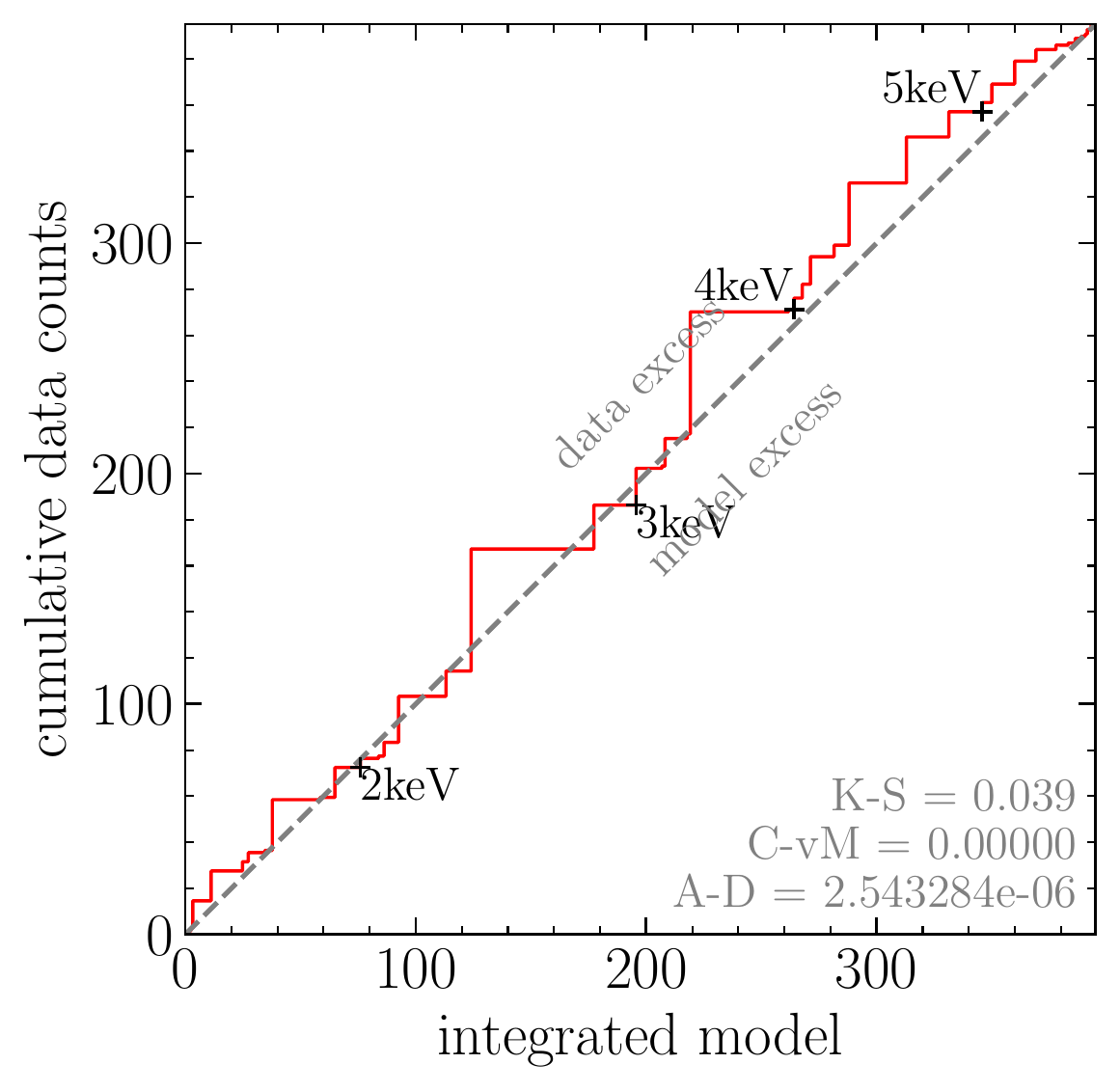}\\
\includegraphics[scale=0.45]{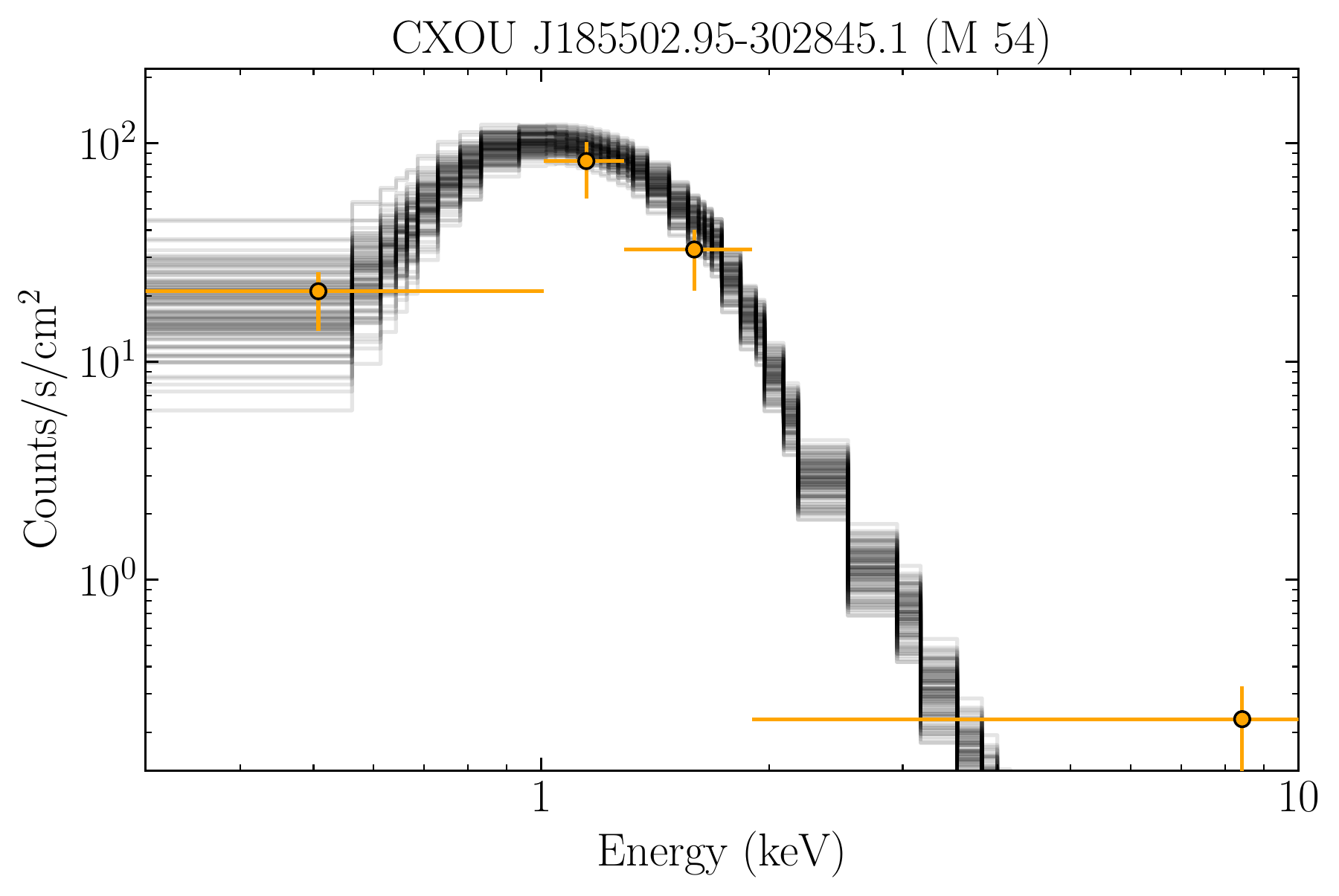}
\includegraphics[scale=0.45]{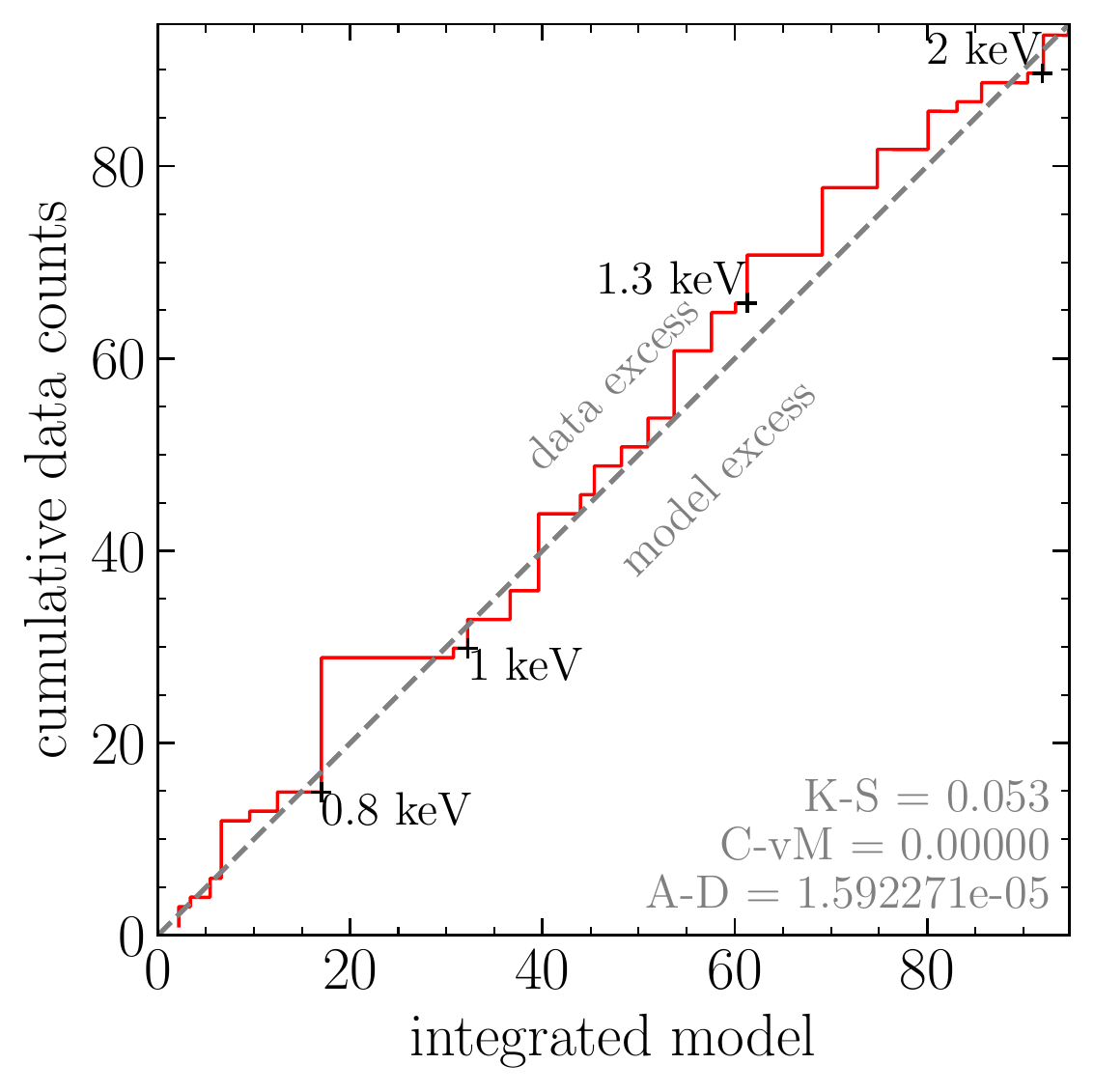}\\
\includegraphics[scale=0.45]{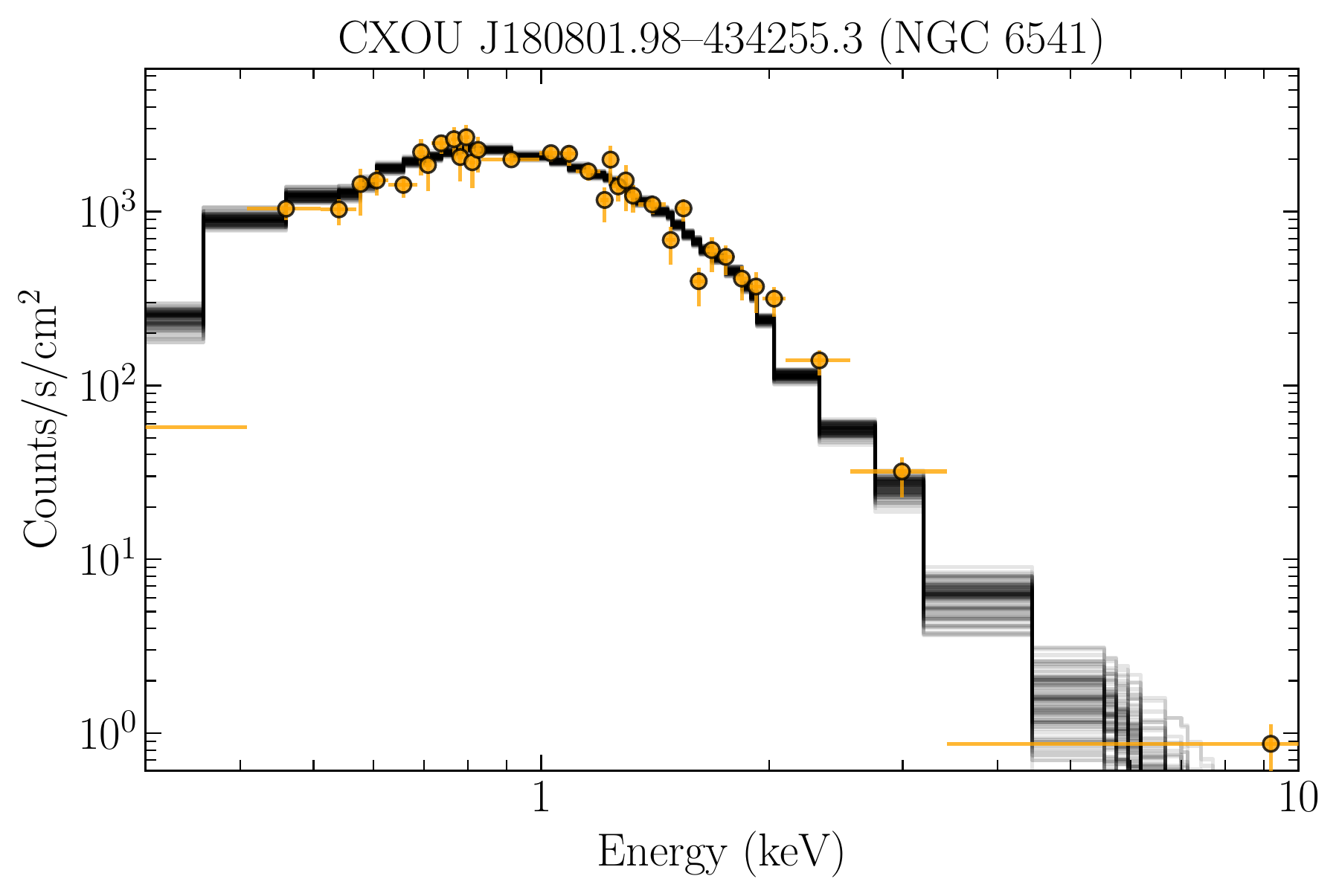}
\includegraphics[scale=0.45]{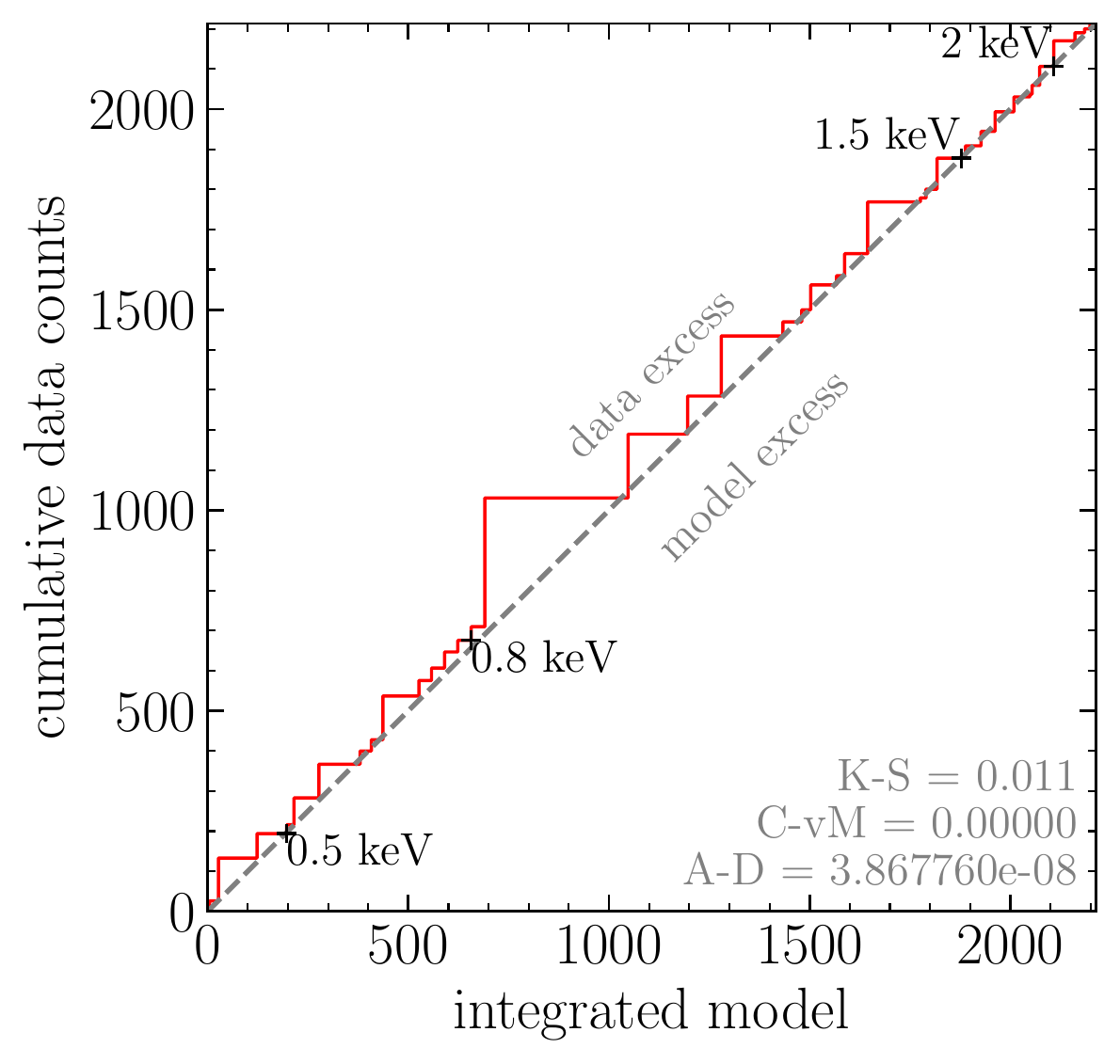}
\caption{X-ray spectra (left) and quantile-quantile (qq) plot for the bright sources analyzed here. The black curves represent a random sample from the posterior distributions for the best-fit model in each case (Table~\ref{tab:brightmodels}). For the purpose of plotting, the data in the spectral plots have been binned adaptively with at least 20 counts. All qq-plots are drawn in the 0.3-10.0 keV band (bottom left corner represents 0.3 keV, top right corner represents 10 keV), where the data were considered for spectral analysis.}
\label{fig:specs1}
\end{figure*}

\begin{figure*}
\centering
\includegraphics[scale=0.45]{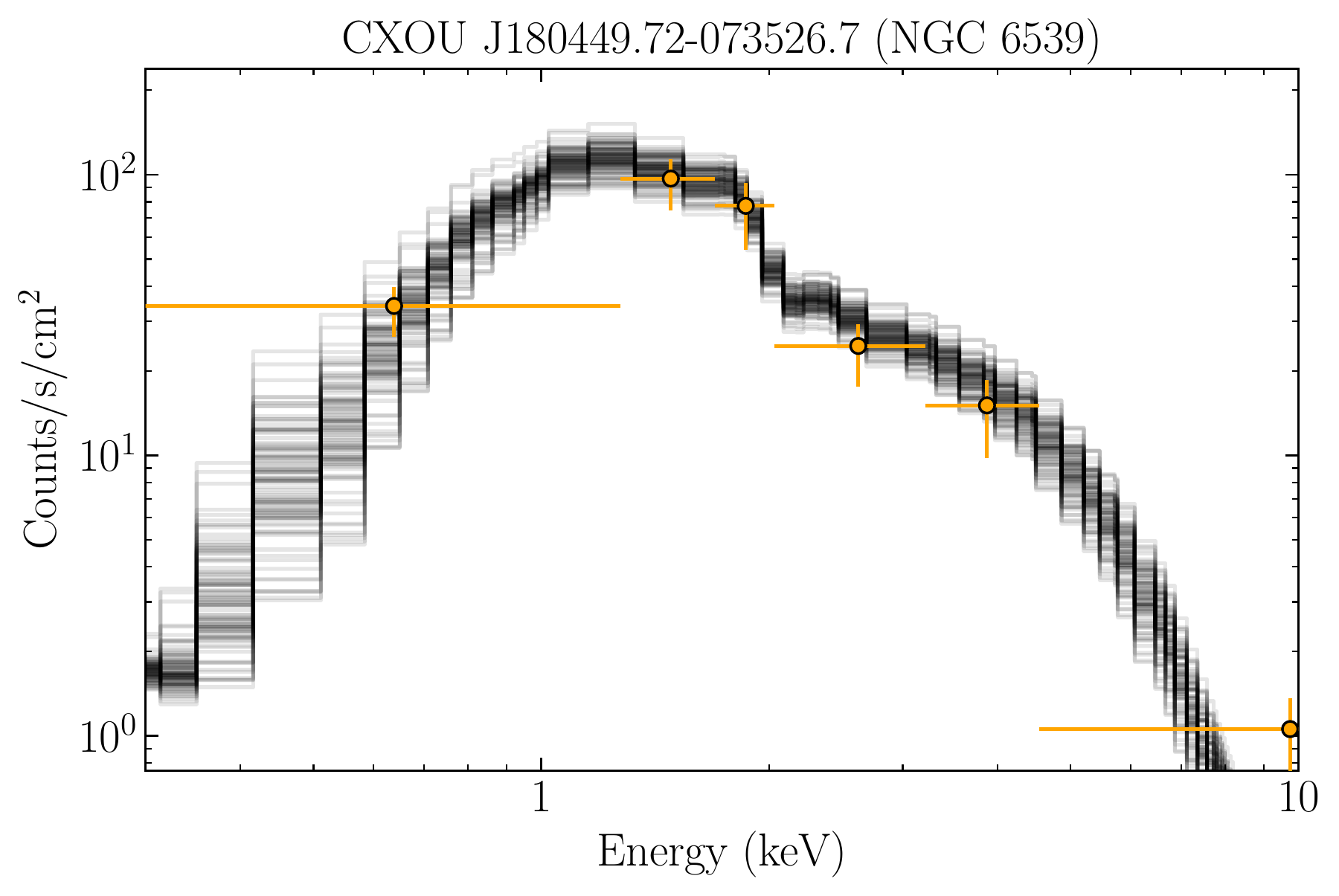}
\includegraphics[scale=0.45]{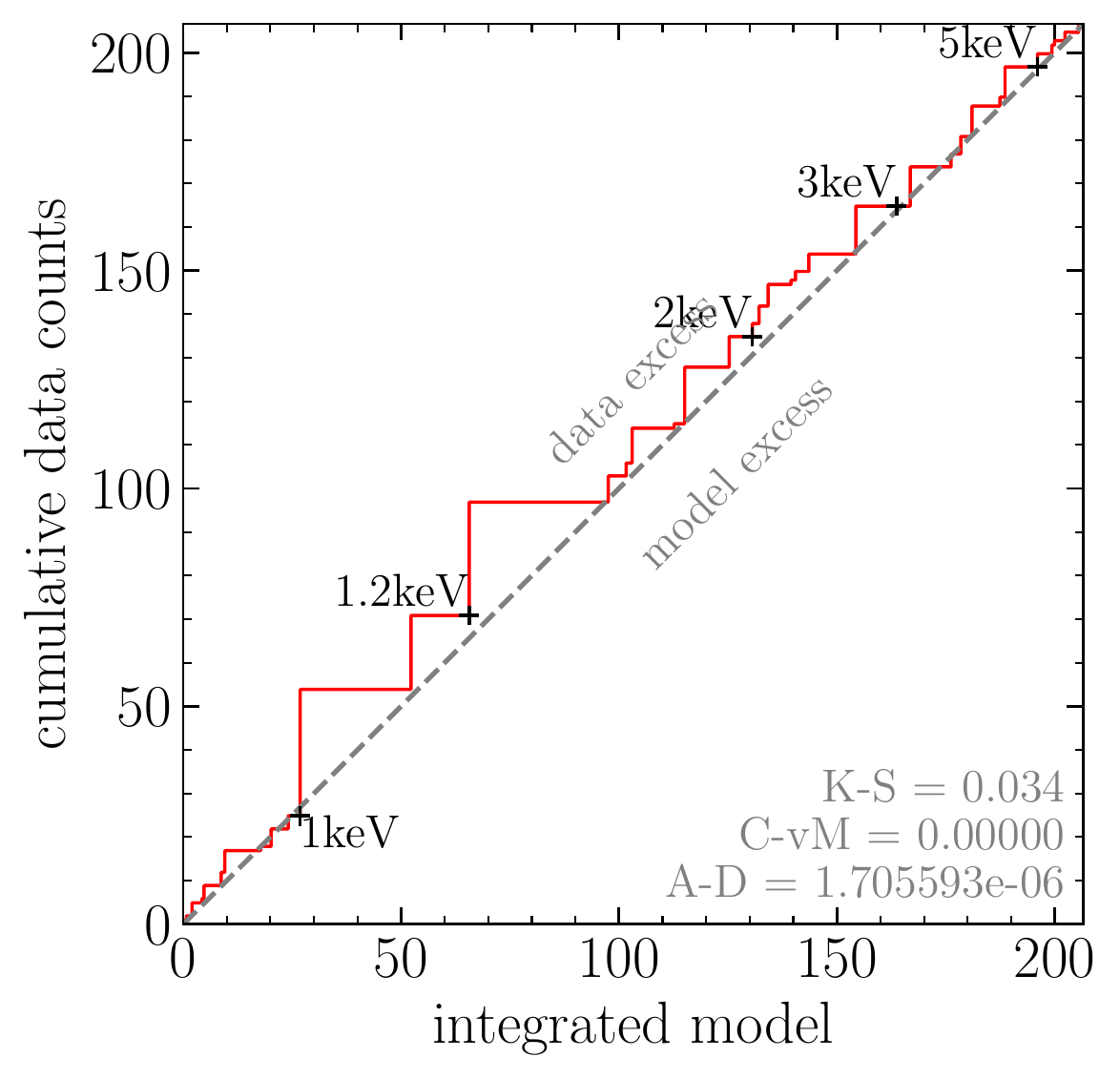}\\
\includegraphics[scale=0.45]{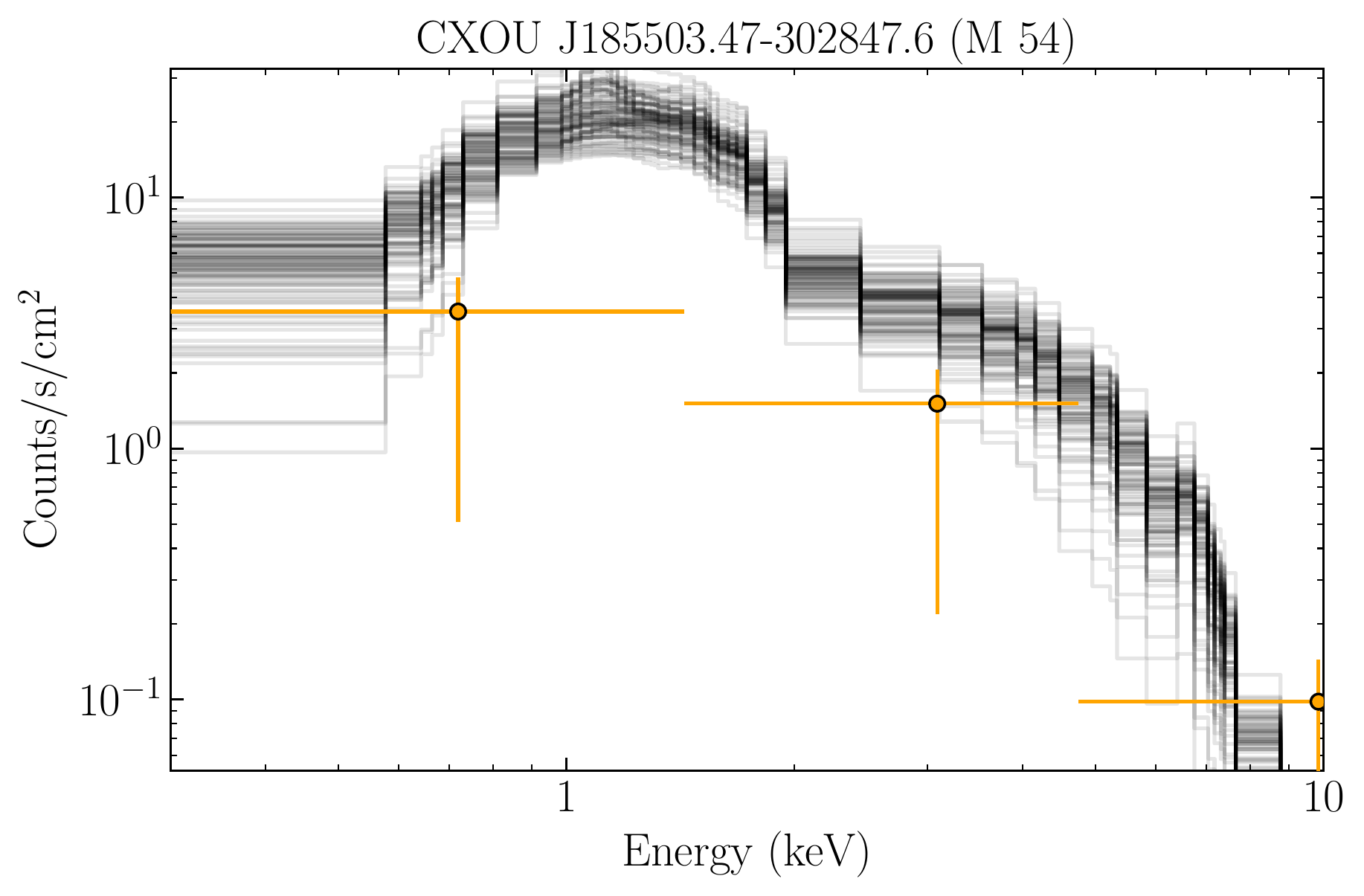}
\includegraphics[scale=0.45]{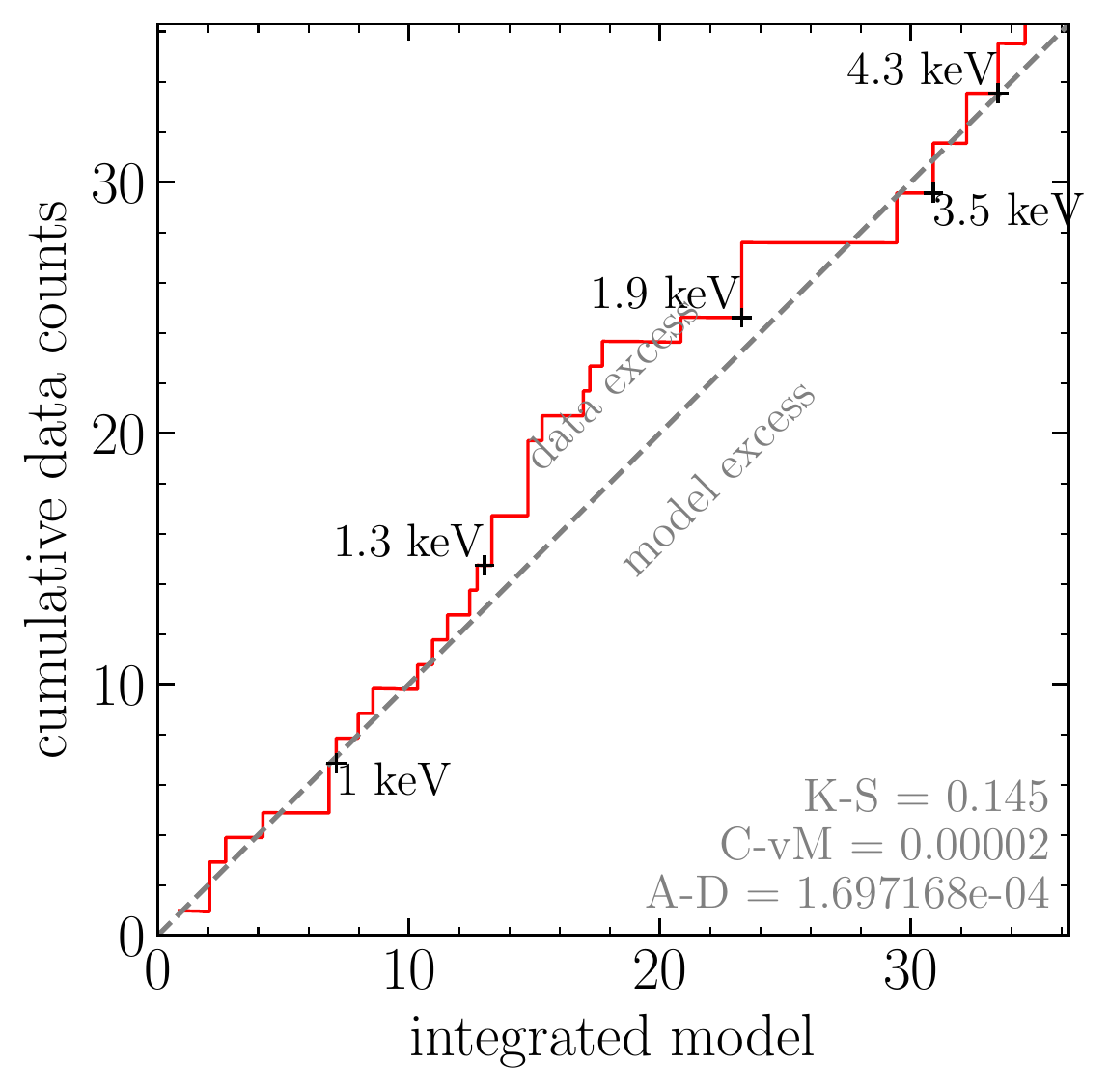}\\
\includegraphics[scale=0.45]{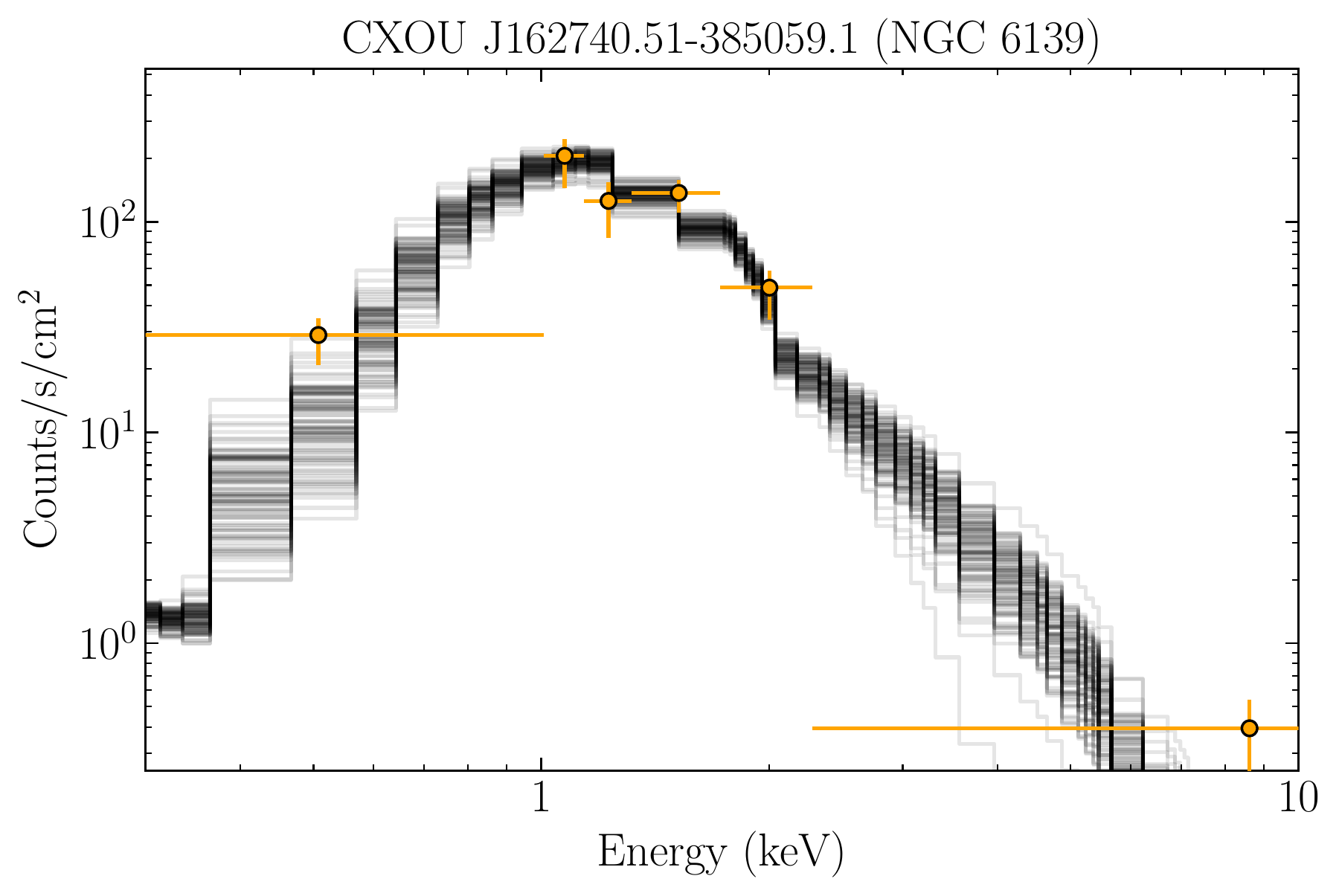}
\includegraphics[scale=0.45]{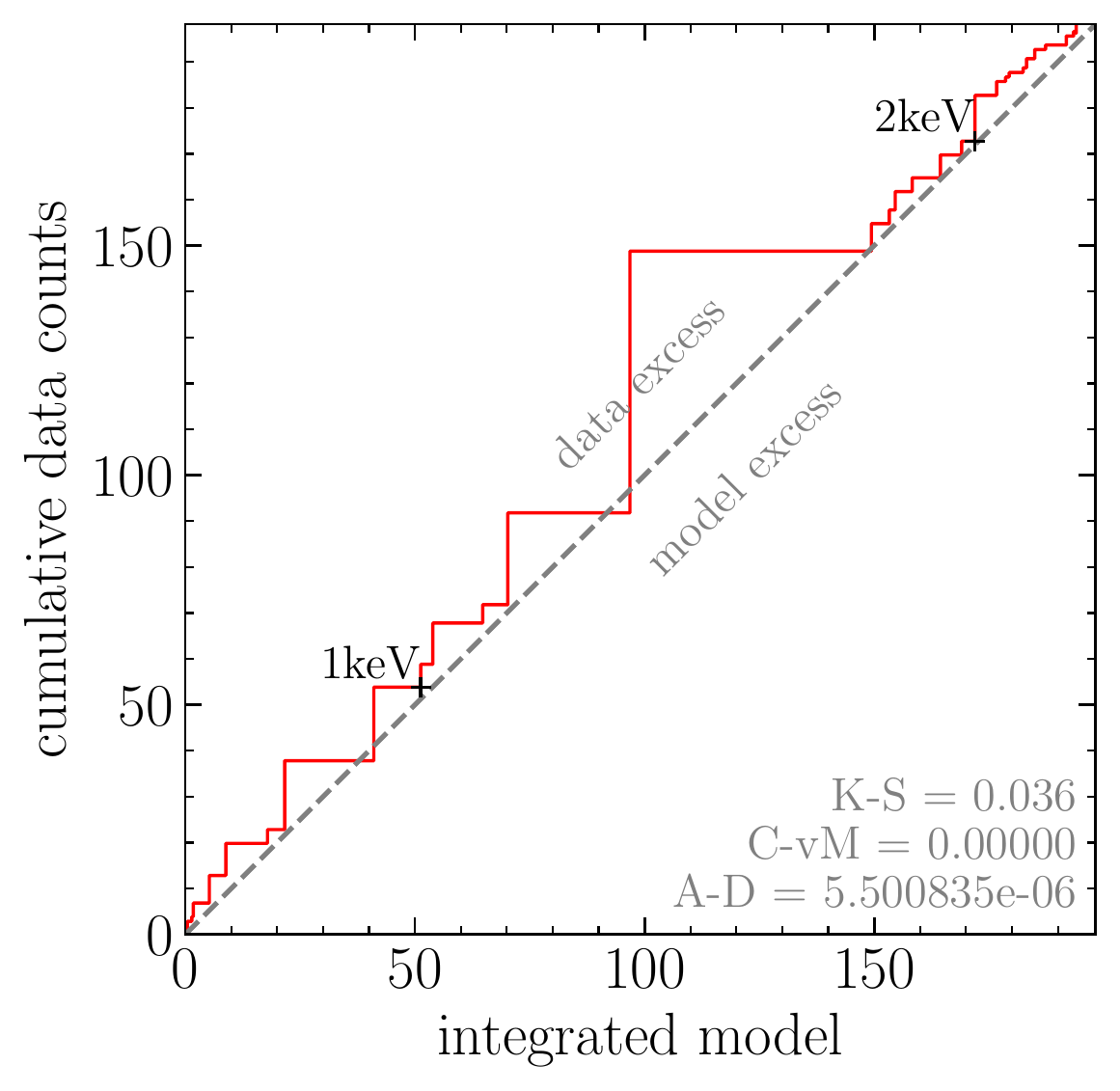}
\caption{Similar to Figure \ref{fig:specs1}. X-ray spectra (left) and quantile-quantile (qq) plot for the bright sources analyzed here. The black curves represent a random sample from the posterior for the best-fit model in each case (Table~\ref{tab:brightmodels}). For the purpose of plotting, the data in the spectral plots have been binned adaptively with at least 20 counts (5 counts in the case of CXOU J185503.47-302847.6). All qq-plots are drawn in the 0.3-10.0 keV band (bottom left corner represents 0.3 keV, top right corner represents 10 keV), where the data were considered for spectral analysis.}
\label{fig:specs2}
\end{figure*}

We use a similar fitting process as in the catalog, taking spectra in the 0.3--10 keV band (binned to contain at least 1 background count per bin), fitting with \textsc{Xspec} and \textsc{bxa}, and using model log-evidence \citep[$\log_{10} Z$;][]{Buchner14} to identify the best model. Besides the power-law and APEC models used in the automated fitting in the catalog, we also tested an absorbed neutron star atmosphere \citep[NSATMOS in Xspec, with neutron star mass and radius fixed to 1.4 M$_\odot$ and 10 km respectively;][]{McClintock04, Heinke06a}  model. When feasible, we considered reasonable multi-component models like NS atmosphere + power-law, or double APEC. These choices allow for a range of physically motivated models, e.g., spectra that might be dominated by the hot surface emission from the NS or a non-thermal accretion flow associated with an NS or BH (or combination thereof), or in the case of an accreting WD, a multi-temperature hot plasma. The spectra and best-fit models for these sources are plotted in Figures~\ref{fig:specs1} and \ref{fig:specs2}, results of model comparisons are tabulated in Table~\ref{tab:brightmodels}, and best-fit values of parameters for each source are reported in Table~\ref{tab:brightspec}. Figures~\ref{fig:specs1} and \ref{fig:specs2} also include quantile-quantile (qq) plots (as produced by \textsc{bxa}) for the accumulated distribution of observed counts versus the expected distribution from the best-fit model (plotted on the left of each qq-plot) to facilitate evaluation of any possible data or model excess.

CXOU J173538.12--303032.0 is a rather bright X-ray source in the field of Terzan~1 (2\farcm5 away from cluster center) with a clear radio counterpart in the ATCA data from the MAVERIC survey (with a flux of 44$\pm9~\mu$Jy at 9 GHz and a $3-\sigma$ upper limit of 47$\mu$Jy at 5.5 GHz). The source shows little evidence of variation within the \acis\ observations. However, there is suggestive evidence for variability among the three observations covering Terzan~1 (KS p-value of 0.01). The X-ray spectrum appears hard, best fit with an absorbed power-law (Table \ref{tab:brightmodels}). Regarding the nature of this source and its association with Terzan~1, it is important to consider that the cluster half-light radius as reported in the Harris catalog (3\farcm8) is likely substantially overestimated \citep[e.g.,][]{Cackett06a}. We did perform the analysis in our catalog assuming the value from the Harris catalog as a conservative upper limit (as removing already-analyzed sources is considerably easier than adding new areas/sources to the analysis). However, when we consider the DSS infrared image of the cluster (Figure~\ref{fig:ter1}), the half-light radius is likely $\lesssim 1'$. In this case CXOU J173538.12--303032.0 is probably unassociated with the cluster, but instead a likely background AGN, as the constraints on the N$_\mathrm{H}$ indicate the source is unlikely to be a foreground object.

CXOU J185502.95--302845.1 is the brightest source in M~54. However, due to the large distance to the cluster (26.5 kpc), the source spectrum contains $<100$ counts, limiting our ability to study the source in detail. There is no evidence for variability of the source in the single \acis\ observation of this GC (KS p-value of 0.65). Fitting the spectrum and comparing the models listed at the beginning of this section, we find an absorbed neutron star atmosphere fits the spectrum best (TBABS$\times$NSATMOS), with an absorbed neutron star atmosphere + power-law slightly less likely (with relative probability of 53\%, Table~\ref{tab:brightmodels}). These probabilities indicate that while a neutron star is highly likely as one of the members of the binary, there is weak evidence for the presence of a second component, such as power-law emission from weak accretion. We found no radio counterpart at the position of the X-ray source in the VLA data from the MAVERIC survey, with a $3\sigma$ upper limit of $< 7.1 \mu$Jy at a frequency of 6.1 GHz, equivalent to a radio limit of $< 3 \times 10^{28}$ \ergs\ at 5 GHz assuming a flat spectrum. This poor upper limit, due to the large distance of M~54, is not very valuable.

CXOU J180801.98--434255.3 is the brightest X-ray source in NGC~6541. The source is bright enough that effects of pile up should be considered in spectral modeling. Thus, we performed a new set of spectral analysis, with the \chandra\ pileup model\footnote{\url{https://heasarc.gsfc.nasa.gov/docs/software/lheasoft/xanadu/xspec/manual/XSmodelPileup.html}} \citep{Davis01} added as a spectral component. We find that the spectrum is best fit with an absorbed NS atmosphere + power-law (PILEUP$\times$TBABS$\times$[NSATMOS+PEGPWRLW]), with an X-ray luminosity of $1.9_{-0.2}^{+0.3}\times10^{33}$ \ergs\ in the 0.5--10 keV band. The model fits the observed spectrum nicely with no significant trend in residuals (Figure~\ref{fig:specs1}, top panels). The power-law component is only loosely constrained, which sometimes suggests it may not be needed. However, comparing model log-evidence between this model and a single absorbed NSATMOS indicates the single-component model is 0.1\% as likely as the two-component model (Table~\ref{tab:brightmodels}). We also find no significant evidence of variability in the single \acis\ observation covering this source. The best-fit X-ray spectral model suggests that this system is likely an NS-LMXB accreting at low levels. Inspecting the ATCA radio data from the MAVERIC survey (Tudor et al., in preparation), we found no detection of a radio source at the position of the X-ray source, with an average $3\sigma$ limit of $< 11.7 \mu$Jy at an average frequency of 7.25 GHz \citep{Tremou18}. Assuming a flat radio spectrum and a cluster distance of 7.5 kpc (based on the Harris catalog), this flux density limit corresponds to a radio luminosity upper limit of $< 3.9 \times 10^{27}$ \ergs\ at the standard reference frequency of 5 GHz. This limit falls close to the radio luminosity observed for accreting BHs at these X-ray luminosities and perhaps slightly above that for transitional MSPs, so in this case the radio data do not strongly constrain the nature of the source.

\begin{figure}
\includegraphics[scale=0.48]{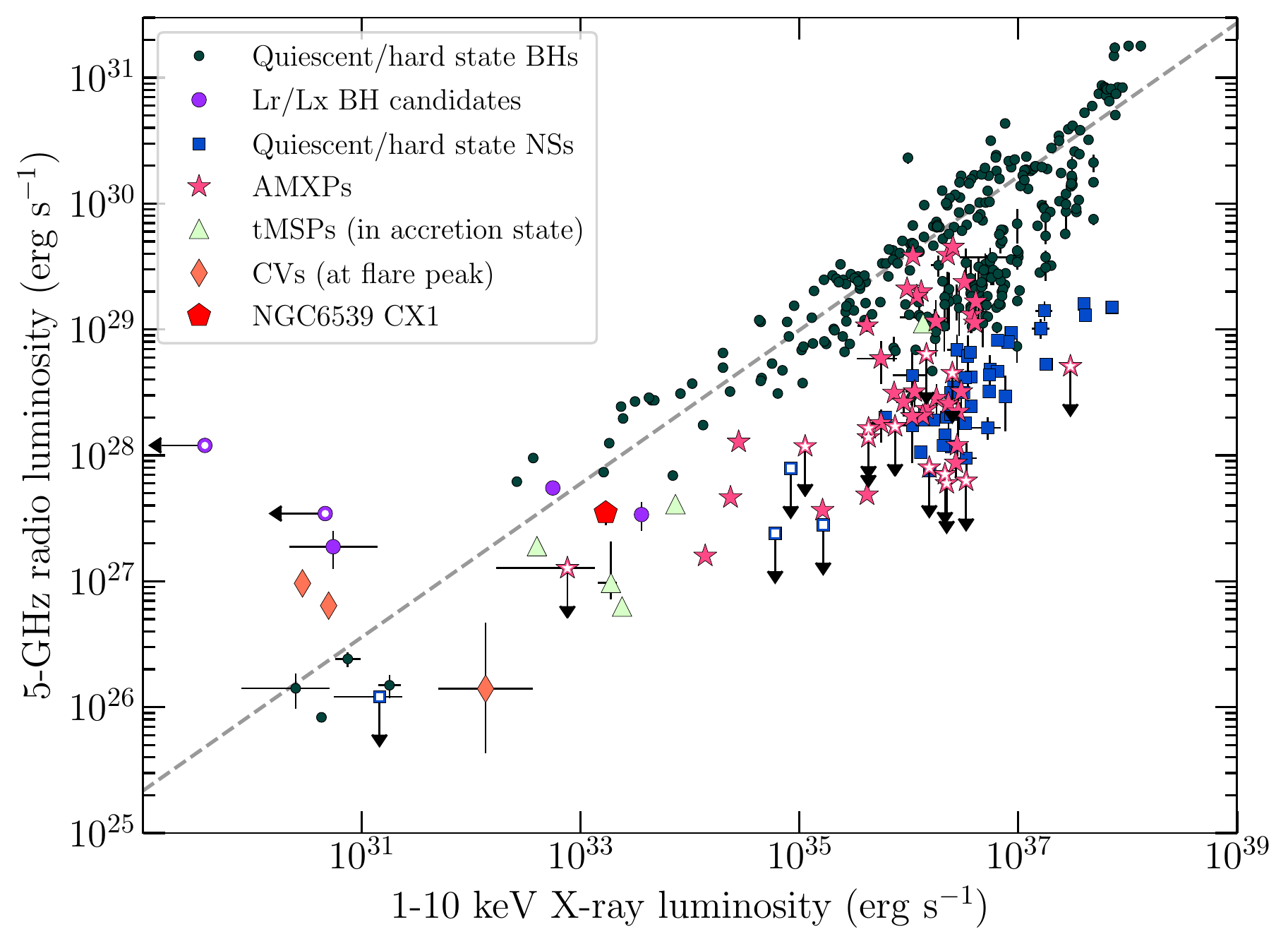}
\caption{Radio/X-ray correlation for accreting compact objects, showing NGC~6539 CX1 (CXOU J162740.51--385059.1) consistent with accreting black holes and tMSPs. However, we note that some accreting neutron stars and accreting millisecond X-ray pulsars may show radio luminosities similar to black holes with similar X-ray luminosity \citep{Russell18,Gusinskaia20}. The dark green circles show known black holes in the Galactic field \citep[e.g.,][]{Miller11, Gallo12, corbel13, Plotkin17}. The dashed gray line shows the best-fit correlation for black holes from \citep{Gallo14}. Purple circles show radio-selected black hole candidates \citep{Strader12, Chomiuk13, Miller-Jones15, TetarenkoB16, Bahramian17a}. Light green triangles show tMSPs \citep{Papitto13, Deller15, Bogdanov18}. Blue squares and pink stars are accreting neutron stars (in the hard state) and accreting millisecond X-ray pulsars \citep{Migliari06, Tudor17}. Orange diamonds are bright CVs and the pulsating white dwarf AR Sco \citep{Russell16, Marsh16}. Plot from \citet{Bahramian18zenodo}.}
\label{fig:lrlx}
\end{figure}

CXOU J180449.72--073526.7 is the brightest X-ray source in NGC~6539, with $L_X \sim 2\times10^{33}$ \ergs. The source shows strong evidence for variability in a 15 ksec \acis\ observation, with a KS p-value of $3\times10^{-7}$. The source shows a hard X-ray spectrum, best fit by an absorbed power-law or a absorbed APEC. An APEC model is 77\% as likely as the power-law model, with neutron star + power-law following with a 55\% relative probabilities. Thus while the presence of a hard power-law/APEC like component is clear, we are unable to infer confidently whether a second (possibly softer) component is needed with the current data. The VLA radio catalog for this cluster from the MAVERIC survey (Shishkovsky et al., in preparation) shows a clear radio continuum counterpart to this X-ray source, with flux densities of $9.6\pm1.9$ and $12.7\pm2.4$ $\mu$Jy at 5.2 and 7.2 GHz, respectively. These values imply a spectral index (for $S_{\nu} \sim \nu^{\alpha}$) of $\alpha=0.51_{-0.76}^{+0.53}$, consistent with a flat or even inverted spectrum. Given the uncertainty in the spectral index, we assume that the 5 GHz flux density is best given by the 5.2 GHz value, corresponding to a luminosity of $3.5\pm0.7 \times 10^{27}$ \ergs. Plotting this source on the radio/X-ray correlation for known LMXBs (Figure~\ref{fig:lrlx}) shows that it is in the region occupied both by BHs and by transitional MSPs in the subluminous disk state. There is mounting evidence that BHs, transitional MSPs, and accreting millisecond pulsars can occupy similar regions of this diagram at $L_X< 10^{35}$ \ergs\ \citep[e.g.,][]{Russell18,Gusinskaia20}, and little is known about the radio behavior of typical NS LMXBs in this regime. Hence other interpretations of this system are possible, and additional follow-up observations will be necessary to better understand this intriguing source.

\begin{figure}
\includegraphics[scale=0.43]{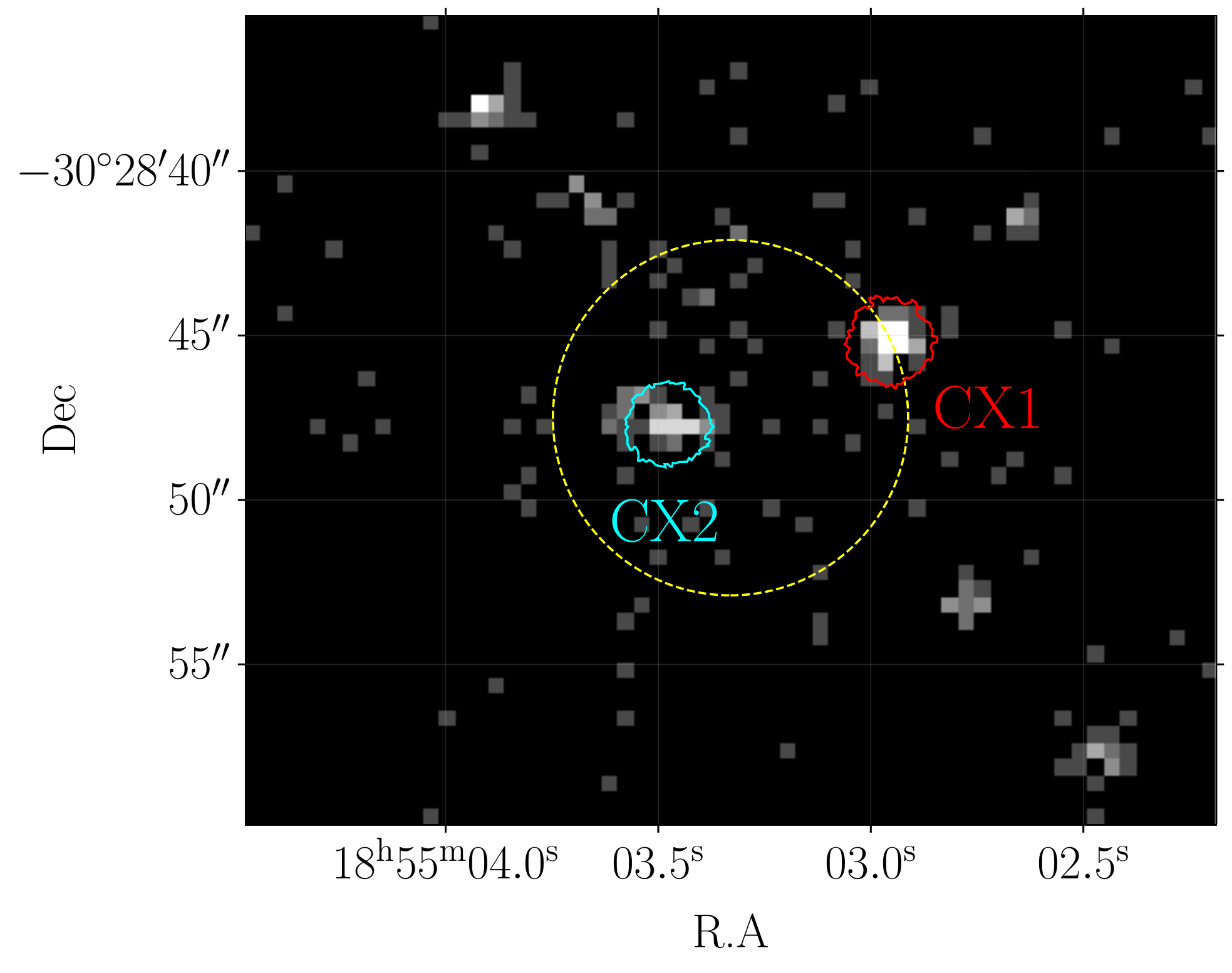}
\caption{Core of M~54 (dashed yellow circle) as seen by \acis. Cluster contains two X-ray sources with $L_X>10^{33}$ \ergs\ as discussed in the text: M~54 CX1 (CXOU J185502.95--302845.1) and M~54 CX2 (CXOU J185503.47--302847.6), denoted by red and cyan regions respectively. The morphology of the M~54 CX2 in the image suggest that it may be confused/blended or extended.}
\label{fig:m54core}
\end{figure}

CXOU J185503.47--302847.6 is the second brightest X-ray source in M~54. The morphology of the source in the image suggest that it may be confused/blended or extended (Figure~\ref{fig:m54core}). Unfortunately, due to the low number of counts ($\sim40$ in the 0.5--10 keV band), it is difficult to disentangle confidently. Assuming it is a single point source, the source spectrum is best fit by an absorbed APEC model. An absorbed power-law or power-law + neutron star atmosphere are also likely models, with probabilities of 15-30\% (relative to APEC). It is difficult to assess the possible nature of this source, due to the possible source confusion and low counts, however the spectrum is consistent with some LMXBs and bright CVs. Similar to CXOU J185502.95--302845.1 (the brightest source in M~54), we found no radio counterpart for this source to a comparable limit.

Lastly, CXOU J162740.51--385059.1 is the brightest source in NGC~6139. This source was previously identified in \rosat\ data \citep{Verbunt01}, but it remains poorly studied. The source shows no significant evidence of variability in the single \acis\ observation of this GC (with a KS p-value of 0.5, indicating no evidence of variability in the current data\footnote{values closer to zero indicate stronger evidence for variability.}). The source clearly shows a rather soft spectrum, best fit by an absorbed NS atmosphere + power-law (TBABS$\times$[NSATMOS+PEGPWRLW]), with an X-ray luminosity of $1.1_{-0.2}^{+0.5}\times10^{33}$ \ergs\ in the 0.5--10 keV band. It is worth noting that absorbed power-law also shows high relative likelihood, and is not entirely ruled out as a possible model (Table~\ref{tab:brightmodels}). These models indicate that the system is probably another weakly accreting NS-LMXB. Considering the ATCA radio data for this source, there again was no evidence of radio emission, with a  $3\sigma$ upper limit of $< 10.8 \mu$Jy at 7.25 GHz. Again assuming a flat spectrum, this corresponds to a reference 5 GHz radio luminosity upper limit of $< 6.6 \times 10^{27}$ \ergs\ at a distance of 10.1 kpc. As for the previous source, this limit does not substantially constrain the nature or state of the accreting compact object.

\begin{deluxetable*}{cccccccc}
\tablecaption{Bright ($L_X>10^{33}$ \ergs) XRBs in our catalog.}
\label{tab:bright}
\tablehead{
\colhead{CXOU J}	 & \colhead{GC}	& \colhead{Alt. Name}       & \colhead{PU?} & \colhead{D$_{\rm{GC}}$}& \colhead{Lx}                 & \colhead{Nature}       & \colhead{Ref.}\\
\colhead{}           & \colhead{}   & \colhead{}                & \colhead{}    & \colhead{(kpc)}        & \colhead{($10^{33}$ \ergs)}  & \colhead{}             & \colhead{}}
\startdata
173617.42--444405.9  &	NGC~6388	& IGR J17361--4441          & Yes           &  9.9	                 & $16~\pm{1}$	                & NS-LMXB?               & 1,2,3         \\
173545.56--302900.0  &	Terzan~1	& Terzan~1 CX1              & No            &  6.7	                 & $2.5~\pm{0.2}$	            & CV (IP)?               & 4,5           \\
180150.32--274923.6  &	Djorg~2	    & OGLE-UCXB-01              & No            &  6.3	                 & $2.4~\pm{0.2}$	            & LMXB? IP?              & 6             \\
173538.12--303032.0  &	Terzan~1	& --                        & No            &  6.7	                 & $2.1~\pm{0.3}$	            & AGN?                   & 7             \\
185502.95--302845.1  &	M~54	    & --                        & No            &  26.5	                 & $2.0~_{-0.3}^{+1.2}$         & NS-LMXB?               & 7             \\
180801.98--434255.3  &	NGC~6541	& --                        & Yes           &  7.5	                 & $1.9~_{-0.2}^{+0.3}$	        & NS-LMXB?               & 7             \\
182432.00--245210.9  &	M~28	    & PSR J1824--2452A          & Yes           &  5.5	                 & $1.83~\pm{0.04}$	            & PSR                    & 8             \\
174805.23--244647.3  &	Terzan~5	& EXO 1745--248             & Yes           &  5.9	                 & $1.79~\pm{0.03}$	            & NS-LMXB                & 9,10,11       \\
182432.49--245208.1  &	M~28	    & IGR J18245--2452          & Yes           &  5.5	                 & $1.70~\pm{0.03}$	            & tMSP                   & 12,13         \\
180449.72--073526.7  &	NGC~6539	& --                        & No            &  7.8	                 & $1.5~\pm{0.2}$	            & BH-LMXB?               & 7             \\
185503.47--302847.6  &	M~54 	    & --                        & No            &  26.5	                 & $1.4~\pm{0.3}$	            & CV?LMXB?               & 7             \\
174805.41--244637.6  &	Terzan~5	& Swift J174805.3--244637   & No            &  5.9	                 & $1.2~\pm{0.1}$	            & NS-LMXB                & 14            \\
162740.51--385059.1  &	NGC~6139	& --                        & No            &  10.1	                 & $1.1~_{-0.2}^{+0.5}$	        & NS-LMXB?               & 7             \\
\enddata
\tablecomments{The ``PU?'' column indicates whether the \chandra\ data on this source may be suffering from pile up. Reported X-ray luminosities are in the 0.5--10 keV band, assuming the source is at the host GC distance. These luminosities are estimated using the best-fit models found in this work. References: 
    1: \citet{Gibaud11},
    2: \citet{Pooley11},
    3: \citet{Bozzo11},
    4: \citet{Wijnands02a}, 
    5: \citet{Cackett06a}, 
    6: \citet{Pietrukowicz19}, 
    7: This work, 
    8: \citet{Bogdanov10b},
    9: \citet{Wijnands05}
    10: \citet{Galloway08},
    11: \citet{RiveraSandoval18a},
    12: \citet{Papitto13},
    13: \citet{Linares14a},
    14: \citet{Bahramian14}.
    }
\end{deluxetable*}

\begin{deluxetable*}{ccccccc}
\tablecaption{Model comparison for the sample of bright sources analyzed in this work.}
\label{tab:brightmodels}
\tablehead{
\colhead{CXOU J}  & \colhead{GC} & \multicolumn{5}{c}{$\log_{10}$Z} \\
\cline{3-7}
\colhead{}        & \colhead{}   & \colhead{AP} & \colhead{2$\times$AP} & \colhead{PL} & \colhead{NS} & \colhead{NS+PL}}
\startdata
173538.12--303032.0 & Terzan~1  &\textbf{-0.58} &  -9.56      &  \textbf{0.00}  & -54.70          & \textbf{-0.11}  \\
185502.95--302845.1 & M~54      &-2.81          &  -6.50      &  -1.17          & \textbf{0.00}   & \textbf{-0.27}  \\
180801.98--434255.3 & NGC~6541  &-63.37         &  -31.35     &  -2.81          & -2.80           & \textbf{0.00}   \\
180449.72--073526.7 & NGC~6539  &\textbf{-0.11} &  -25.03     &  \textbf{0.00}  & -52.26          & \textbf{-0.26}  \\
185503.47--302847.6 & M~54      &\textbf{0.00}  &  -14.48     &  \textbf{-0.53} & -9.70           & \textbf{-0.37}  \\
162740.51--385059.1 & NGC~6139  &-2.48          &  -8.29      &  \textbf{-0.20} & -1.92           & \textbf{0.00}   \\
\enddata
\tablecomments{We used model evidence ($\log_{10}$Z) as implemented in BXA, normalized to the model with the highest evidence in each case. In this representation, evidence indicates relative likelihood of the models. Models tested here are APEC (AP), power-law (PL), neutron star atmosphere (NS), and reasonable combinations of them. Values in bold represent the most likely models.}
\end{deluxetable*}

\begin{deluxetable*}{cccccccc}
\tablecaption{Best-fit values of parameters and uncertainties based on posterior distributions (median and 90\% credible intervals) for the most likely models.}
\label{tab:brightspec}
\tablehead{
\colhead{CXOU J}    & \colhead{GC} & \colhead{Model}& \colhead{N$_\mathrm{H}$}                 & \colhead{kT}  & \colhead{$\Gamma$} & \colhead{Unabs. Flux}& \colhead{$\log_{10}$(A--D)} \\
\colhead{}          & \colhead{}   & \colhead{}     & \colhead{($10^{21}$ cm$^{-2}$)} &\colhead{(keV)}& \colhead{}         & \colhead{($10^{-14}$\ergcms)}   & \colhead{}
}
\startdata
173538.12--303032.0 & Terzan~1  & PL          & $35_{-7}^{+8}      $ & $ -                          $ & $ 1.2\pm0.3         $ & $ 39_{-3}^{+5}     $ & -5.6  \\
185502.95--302845.1 & M~54      & NS          & $2\pm1             $ & $ 0.14\pm0.006               $ & $ -                 $ & $ 2.4_{-0.4}^{+1.5}$ & -4.8  \\
180801.98--434255.3 & NGC~6541  & NS+PL       & $1.8_{-0.4}^{+1.5} $ & $ 0.133_{-0.006}^{+0.003}    $ & $ 2.9_{-2.2}^{+1.0} $ & $ 29_{-3}^{+6}     $ & -7.4  \\
180449.72--073526.7 & NGC~6539  & PL          & $5\pm2             $ & $ -                          $ & $ 1.6\pm0.3         $ & $ 21\pm3           $ & -5.8  \\
185503.47--302847.6 & M~54      & AP          & $<1.4              $ & $ 11_{-6}^{+13}              $ & $ -                 $ & $ 1.7\pm0.4        $ & -3.8  \\
162740.51--385059.1 & NGC~6139  & NS+PL       & $9_{-2}^{+3}       $ & $ 0.08\pm0.05                $ & $ 3.8_{-1.3}^{+0.6} $ & $ 9_{-2}^{+4}      $ & -5.3  \\
\enddata
\tablecomments{kT respresents the temperature (neutron star atmosphere, or hot plasma in APEC). $\Gamma$ is power-law photon index. Flux values are unabsorbed in the 0.5--10 keV band. $\log_{10}$(A--D) represents Anderson-Darling statistics value (not p-values). Smaller values indicate a better fit.}
\end{deluxetable*}

\subsection{Populations of moderately bright X-ray sources}

We can use the results from our survey of bright X-ray sources in 38 globular clusters to place interesting constraints on several X-ray source populations. 

\subsubsection{Intermediate polars}
First, we discuss X-ray bright intermediate polars (IPs). Intermediate polars are unusual among CVs in being highly efficient at converting accretion energy into X-rays even at high accretion rates, unlike disk-accreting CVs \citep{Patterson85,Patterson94}.   \citet{Pretorius14} showed that there are two populations of IPs, an X-ray bright sample, and a low-luminosity sample. With Gaia distances, it has become clear that the bright IPs generally have X-ray luminosities of  $10^{33}$--$2\times10^{34}$ erg/s \citep{Suleimanov19,Schwope18}, while the fainter IP population, mostly shorter-period systems below the period gap, have $L_X$ between $10^{30}$--$10^{32}$ erg/s. Many bright IPs have been detected in the Galactic Field with all-sky hard X-ray coded-mask surveys \citep[INTEGRAL and Swift/BAT;][]{Barlow06,Baumgartner13}. Of particular importance is that the same bright IP systems that can be surveyed over the whole sky are also all bright enough to appear in our $L_X>10^{33}$ erg/s sample.

\citet{Pretorius14} carefully estimated the space density of X-ray bright IPs using the Swift/BAT survey, finding a local space density of $1^{+1}_{-0.5}\times10^{-7}$ pc$^{-3}$ (much of the uncertainty comes from the uncertain disk scale height of the IP population). \citet{Schwope18} used Gaia DR2 distances to update the space density to $7.4^{+4.8}_{-1.7}\times10^{-8}$ pc$^{-3}$. For a local stellar density of 0.1 stars/pc$^3$ \citep{Gliese91}, that equates to $7\times10^{-7}$ bright IPs per star, or (assuming an average stellar mass of 0.5 \Msun) $1.5^{+1.0}_{-0.3}\times10^{-6}$ bright IPs per \Msun. 
  
We sum the masses of the 38 globular clusters in our sample, using the calculations of \citet{Baumgardt18}\footnote{\url{https://people.smp.uq.edu.au/HolgerBaumgardt/globular/parameter.html}}, giving a total mass of $1.37\times10^7$ \Msun, for a predicted number of $21^{+13}_{-5}$ bright IPs in these globular clusters, assuming the field rate of IPs per stellar mass. 
  
However, we see only two (maybe three) plausible IP candidates among our moderately bright globular cluster X-ray sources. This is a factor of 10 fewer IPs than expected, at more than 3$\sigma$ significance. Intriguingly, this is the opposite result from \citet{Grindlay95}, who suggested an overabundance of magnetic CVs in globular clusters; several papers have since attempted to explain this overabundance  \citep[e.g.][]{Dobrotka06,Ivanova06}.

It is not obvious how to explain the opposite result of fewer IPs in globular clusters. The destruction of the wide binaries that are CV progenitors (pre-common envelope) in the dense environments of globular clusters is a logical direction. However, the reduction of CV numbers in globular clusters has been estimated at only a factor of 2--3 \citep{Shara06,Haggard09}, and CV numbers are higher in denser globular clusters \citep[e.g.][]{Pooley03,Heinke20}. Belloni et al. (in prep) will discuss possible solutions to this problem.

\subsubsection{Symbiotic stars}
We have no good candidate symbiotic stars in our $L_X>10^{33}$\,\ergs\ sample.  In the field, symbiotic stars with  $L_X\sim10^{33}$\,\ergs\ but without strong optical emission lines (suggesting an accretion luminosity below $\sim10^{34}$\,\ergs; \citealt{Mukai16}) appear to be somewhat common \citep{vandenBerg06,Hynes14,Mukai16}, though not well-studied. \citet{Lu06} estimate 1,200--15,000 symbiotic stars in the galaxy with accretion luminosity above 10 \Lsun. Systems with lower accretion rates \citep[as in][]{Mukai16} should be more numerous. \citet{Munari92} empirically estimate $3\times10^5$ symbiotic stars in the galaxy, but a theoretical prediction of the numbers of symbiotics with lower accretion luminosities would be of great interest.

If we take $3\times10^5$ symbiotic stars in our Galaxy (assumed $10^{11}$ \Msun), we can estimate that $\sim$40 symbiotic stars should be present in our globular cluster sample. However, their luminosity function is unknown. If we make the (doubtful) assumptions that \citet{vandenBerg06} detected all the symbiotics in their Galactic Bulge fields of view, that they are all symbiotics, and that those systems are all located at 8 kpc, then we may infer that 1/13 symbiotic stars have $L_X\sim10^{33}$ erg/s. This would suggest $\sim$3 symbiotic stars in our sample. The lack of symbiotic stars (if the assumptions above prove reasonable) should not be surprising, however, due to the destruction of wide binaries in globular clusters \citep[e.g.,][]{Belloni20b}. 

There have been suggestions of symbiotic stars in some globular clusters. \citet{Henleywillis18} identified a  symbiotic star in $\omega$ Cen, through a robust identification of an  $L_X\sim3\times10^{32}$\,\ergs\ X-ray source with a red giant (a carbon star). \citet{Belloni20b} suggest that the symbiotic star in $\omega$ Cen has been  mis-classified, as they do not see an optical H$\alpha$ emission line in their SALT spectrum. A requirement of strong optical H emission lines has often been used to confirm symbiotic stars \citep[see, e.g.][]{Allen84,Belczynski00}. However, a number of symbiotic systems have been suggested based on other wavelengths, such as X-ray emission, without H emission lines \citep{Hynes14,Bahramian14}.  \citet{Mukai16} confidently identify SU Lyncis as a white dwarf accreting from a red giant, though it shows only weak H lines. \citet{Munari19} show that even weak H emission lines are not always present in high-resolution spectra of SU Lyncis (see also \citealt{Kenyon16} for occasional absence of H emission lines in EG And), and propose an alternative definition of symbiotic stars in which any binary where a WD or NS accretes enough material from a RG to be detected at any wavelength is considered a symbiotic binary. Thus careful multi-wavelength observations are needed for accurate classification of symbiotic systems.

\subsubsection{Transitional millisecond pulsars}
We can also constrain the numbers of tMSPs in their ``active'' state (with $L_X\sim10^{33}-10^{34}$\,\ergs).  Note that it is currently unclear whether the ``active'' state represents active, low-level accretion onto the NS \citep[e.g.][]{Jaodand16}, or  enhanced radio pulsar emission \citep{Ambrosino17}.

We have 1-2 ``active'' tMSP candidates among our 38 clusters (IGR J18245-2452 and perhaps Ter5 CX1, \citealt{Bahramian18}). We can roughly estimate the total number of qLMXBs in these clusters by adding up the calculated stellar interaction rates for these clusters (tabulated in \citealt{Bahramian13}), which amount to 40\% of the total stellar interaction rate of the Milky Way's globular cluster system. Estimating 250 qLMXBs in the entire globular cluster system \citep[extrapolating from the 5 qLMXBs in 47 Tuc][]{Heinke05a,Heinke05b}, this implies $\sim$100 quiescent LMXBs among these clusters. 

From these numbers, we can conclude that the average quiescent LMXB spends only $\sim$2\% of its lifetime in a transitional MSP ``active'' state. Of course, we do not know what fraction of LMXBs pass through a transitional MSP phase, or for how long, nor do we know what fraction of the transitional MSP phase these systems spend in their ``active'' state; our analysis cannot separate these questions. 

A caveat is that we will miss tMSPs in the ``active'' state if we only use stacked data in which a tMSP is only ``active'' for a short time. For instance, Ter 5 CX1 was rarely X-ray bright, so that  its averaged $L_X$ in the stacked image falls below $10^{33}$\,\ergs. However, only 3 of our clusters were observed at more than 2 epochs, suggesting that we have probably found the ``active'' tMSPs in our sample. Note that the 2 tMSPs we have found were in the 3 clusters observed the most, suggesting that a number of tMSPs may be hidden in our clusters \citep[e.g. 47 Tuc V,][]{Ridolfi16}.

\section{Conclusion}\label{sec:conc}
In this paper, we provide a deep and extended catalog of faint ($L_X<10^{35}$ \ergs) X-ray sources in a sample of 38 Galactic GCs. We perform photometry, variability, and spectral analysis for each of more than 1100 sources and catalog the results of these analyses. We also investigate the X-ray properties, nature and population of faint XRBs in GCs and their luminosity function, which is now the deepest X-ray luminosity function of GC XRBs.

The X-ray sources reviewed in \S\ref{sec:brightknown} and the new ones identified in \S\ref{sec:brightnew} provide us with a large (and fairly complete) sample of XRBs with $10^{33} \leq L_X \leq 10^{35}$ \ergs\ in this set of Galactic GCs. This sample allows us to explore the population of weakly accreting systems in GCs. While such sources in the field are heterogeneous, containing BH-LMXBs and NS-LMXBs, transitional MSPs, CVs, and symbiotic X-ray binaries \citep[e.g.,][]{Shaw20}, the GC sample here seems somewhat less diverse, with a large population of NS-LMXBs, two CV candidates and a BH-LMXB/tMSP candidate.

Besides the systems reviewed and identified in this work with $10^{33} \leq L_X \leq 10^{35}$ \ergs, there are three other GC LMXBs identified in Galactic GCs with similar luminosities: the BH-LMXB candidate 47 Tuc X9 \citep{Miller-Jones15,Bahramian17a},  the candidate NS-LMXB M~15 X-3 \citep{Heinke09b, Arnason15}, and the unusual variable source NGC~6652B \citep{Stacey12} (these GCs were not in our current paper, see \S\ref{sec:sample}). It is worth noting that this sample is not complete. It is possible for some sources in our survey which show $L_X <10^{33}$ \ergs\ to have heavy partial covering/intrinsic absorption, and thus for their true luminosity to be substantially higher. There are also likely to be similar systems in GCs not covered in our survey or in  other studies. In the $L_X > 10^{35}$ \ergs\ regime, out of 21 transient and persistent systems identified in Galactic GCs so far, 20 are confirmed or candidate NS-LMXBs, with one of them  a confirmed tMSP \citep[IGR J18245--2452,][]{Papitto13}. The only exception so far has been the transient source 1E 1339.8+2837 in M~3, which appears to be an accreting white dwarf that showed an episode of (somewhat underluminous) supersoft X-ray emission \citep{Grindlay93, Verbunt95, Edmonds04}. In the $10^{33}$ -- $10^{35}$ \ergs\ range, while NS-LMXBs appear to still hold a majority, the contributions from other classes of objects become noticeable. Out of 15 systems identified so far (12 reviewed or identified in this study, plus the 3 systems mentioned above), 10 are candidate or confirmed NS-LMXBs (with one plausible tMSP candidate), 2 are candidate BH-LMXBs, 2 are candidate CVs, and one is an unusually X-ray bright pulsar. The classification of numerous, fainter sources with $L_X < 10^{33}$\,\ergs\ will be an essential focus of future work.

Lastly, we inspected the populations of classes of bright $10^{33}$ -- $10^{35}$ \ergs\ sources in our sample. We noticed a significant deficit of bright ($L_X>10^{33}$\,\ergs) intermediate polar CVs in GCs, compared to their abundance in the field. This is intriguing, as GCs have long been thought to be {\it over-abundant} in magnetic CVs, particularly intermediate polars. We find (more speculative) evidence for an under-abundance of symbiotics in GCs, which is not surprising considering their likelihood of disruption by encounters, due to their large orbits. We also show that quiescent LMXBs in GCs may spend $\sim$2\% of their lifetimes in a transitional MSP ``active'' state, with $L_X\sim10^{33}-10^{34}$\,\ergs\ (we do not constrain their lifetimes in the lower-$L_X$ ``passive'' transitional MSP state). 

\acknowledgments
We thank the anonymous referee for their helpful comments. AB thanks Patrick Broos, Diogo Belloni, Liliana Rivera Sandoval and Scott Ransom for helpful discussions. JCAMJ is the recipient of an Australian Research Council Future Fellowship (FT140101082) funded by the Australian government. LC is grateful for the support of NASA grants Chandra-G06-17040X, Chandra-GO7-18032A, HST-GO-14351. JS acknowledges support from a Packard Fellowship, NASA grants Chandra-GO3-14029X, Chandra-GO5-16036X, and Chandra-GO8-19122X, and NSF grants AST-1308124 and AST-1514763. COH \& GRS acknowledge NSERC Discovery Grants RGPIN-2016-04602 and RGPIN-2016-06569 respectively, and COH also a Discovery Accelerator Supplement. KLL is supported by the Ministry of Science and Technology of the Republic of China (Taiwan) through grant 108-2112-M-007-025-MY3. We acknowledge extensive use of NASA's Astrophysics Data System Bibliographic Services, Arxiv, and SIMBAD \citep{Wenger00}.

%

\vspace{5mm}
\facilities{\chandra(ACIS), VLA, ATCA}


\software{ACIS-Extract \citep{Broos10}, 
          Astropy \citep{Astropy13,Astropy18}, 
          BXA \citep{Buchner14}, 
          CIAO \citep{Fruscione06},
          Corner \citep{ForemanMackey16},
          HEASOFT \citep{HEASARC14}, 
          IPython \citep{Perez07}, 
          Jupyter \citep{Kluyver16}, 
          Matplotlib \citep{Hunter07}, 
          MultiNest \citep{Feroz19},
          Numpy \citep{oliphant06,vanderwal011}
          Pwdetect \citep{Damiani97},
          SAOImage DS9 \citep{Joye03}, 
          Scipy \citep{Virtanen20}, 
          XSPEC \citep{Arnaud96}.
          }


\clearpage
\appendix

\section{Overview of catalog columns}\label{sec:columns}
The final published catalog online contains 123 columns, including details of source coverage and properties. These detailed data are provided to allow careful further analysis by readers. The catalog in its complete form is available in the electronic version of this article. Here, in Table \ref{tab:columns}, we provide detailed description of all the columns in the catalog.

\begin{longrotatetable}
\begin{deluxetable*}{lS}
\tablecaption{}
\label{tab:columns}
\tablehead{\colhead{Column name}    & \colhead{Column Description}}
\startdata
CXOU\_J & Source designation following the \chandra\ convention for naming unregistered sources. \\
detection\_quality\_flag & If a source has a minimum false probability value of $\geq 1\%$, we classify it as a poor detection (with a detection quality flag value of 2). If a source has a minimum false probability value of $< 1\%$ and a net source count $< 5$ (in the 0.5--10 keV band), we classify it as a marginal detection (detection quality flag value = 1). Finally, if a source has a minimum false probability value of $< 1\%$ and a net source count $\geq 5$, we classify it as a confident detection (detection quality flag value = 0). For more details see \S\ref{sec:validity} and Figure~\ref{fig:detection}. \\
spectrum\_quality\_flag & Indicating spectrum quality based on total number of source counts: if a source has $\gsim100$ counts, it would have relatively reliable spectral analysis (and we assign a flag value of 0). If the total number of counts is somewhere between $\lsim100$ and $\gsim20$, the estimates are less reliable and should be taken with caution (flag value of 1). Lastly, if a source has $\lsim20$ counts, spectral analysis is merely suggestive, and model comparison is not to be taken with confidence (flag value of 2). \\
pile\_up\_flag & Whether the source is in danger of being piled up. If True, photometry and spectroscopy presented in this catalog are likely inaccurate. Only 15 sources are identified as potentially piled in our catalog. \\
foreground\_flag & True means that the source could be a foreground object. This is determined based on comparing the upper limit on the source N$_\mathrm{H}$ (from the best-fit model) compared to the host cluster N$_\mathrm{H}$ (estimated based on E(B-V)). False doesn't necessarily reject that source is a foreground object. Just that the source N$_\mathrm{H}$ is not inconsistent with the host cluster. Refer to \S\ref{sec:flags} and \ref{sec:caveats} for details on this assessment and its caveats. \\
Prob\_AGN & Probability that the source may be a background AGN. Estimated based on source flux, population of sources in the cluster and the location of the source in the cluster. If the source is located outside the GC half-light radius, this probability is not estimated. See \S\ref{sec:flags} for details. \\
abs\_astrometry\_flag & If the coordinates are corrected for absolute astrometry. This is currently only done for Terzan~5. For the rest of the catlaog, one should consider an additional 0.8'' uncertainty to consider \chandra's astrometry accuracy. \\
RA\_centroid & Source RA estimated via centroiding (Coordinates used for most analyses in this paper). \\
Dec\_centroid & Source Dec estimated via centroiding (Coordinates used for most analyses in this paper). \\
Centroid\_error & Centroiding uncertainty radius (arcsec). \\
RA\_recon & Source RA estimated via image reconstruction. \\
Dec\_recon & Source Dec estimated via image reconstruction. \\
RA\_correlate & Source RA estimated via correlating image and PSF. \\
Dec\_correlate & Source Dec estimated via correlating image and PSF. \\
Num\_obs\_total & Total number of observations covering the source. \\
Num\_obs\_merged & Number of observations merged for this source to enhance source validity. \\
Exposure & Total source-specific exposure time from the merged observations (s). \\
Primary\_ccd & The \acis\ CCD (ACIS-I or ACIS-S) which contain most/all of the observations covering the source. \\
theta\_avg & Average off-axis angle (arcmin). \\
theta\_low & Minimum off-axis angle (arcmin). \\
theta\_high & Maximum off-axis angle (arcmin). \\
psf\_fraction & Average PSF fraction. \\
src\_cnts\_* & Total source counts in different bands. \\
bkg\_cnts\_* & Total background counts in different bands. \\
backscale\_* & Background scaling factor in different bands. \\
net\_cnts\_* & Net source counts in different bands. \\
net\_cnts\_sigma\_up\_* & 1-sigma upper-limit on net source counts in different bands. \\
net\_cnts\_sigma\_low\_* & 1-sigma lower-limit on net source counts in different bands. \\
significance\_* & Photometric signal to noise ratio in different bands. \\
prob\_no\_source\_* & p-value for no-source hypothesis in different bands. \\
min\_KS\_single\_obs & Minimum p-value for KS variability test within each observation. These values should be considered with caution. See \S\ref{sec:phot/var} and \ref{sec:caveats} for further details. \\
KS\_all\_data & p-value for KS variability test for all observations. These values should be considered with caution. See \S\ref{sec:phot/var} and \ref{sec:caveats} for further details. \\
ae\_abs\_flx\_*kev & Absorbed flux in different bands based on a power-law fit in AE (\ergcms). \\
ae\_unabs\_flx\_*kev & Unabsorbed flux in different bands based on a power-law fit in AE (\ergcms). \\
fit1\_pl\_nh & BXA Power-law fit N$_\mathrm{H}$ (cm$^{-2}$). \\
fit1\_pl\_nh\_er- & BXA Power-law fit lower uncertainty (cm$^{-2}$). \\
fit1\_pl\_nh\_er+ & BXA Power-law fit upper uncertainty (cm$^{-2}$). \\
fit1\_pl\_gamma & BXA Power-law fit photon index. \\
fit1\_pl\_gamma\_er- & BXA Power-law fit photon index lower uncertainty. \\
fit1\_pl\_gamma\_er+ & BXA Power-law fit photon index upper uncertainty. \\
fit1\_pl\_flx\_0.5-10 & BXA Power-law fit unabsorbed flux in the 0.5-10 keV band (\ergcms). \\
fit1\_pl\_flx\_0.5-10\_er- & BXA Power-law fit unabsorbed flux lower uncertainty in the 0.5-10 keV band (\ergcms). \\
fit1\_pl\_flx\_0.5-10\_er+ & BXA Power-law fit unabsorbed flux upper uncertainty in the 0.5-10 keV band (\ergcms). \\
fit1\_pl\_lum\_0.5-10 & BXA Power-law fit Luminosity in the 0.5-10 keV band (assuming cluster distance; \ergs). \\
fit1\_pl\_lum\_0.5-10\_er- & BXA Power-law fit Luminosity lower uncertainty in the 0.5-10 keV band (\ergs). \\
fit1\_pl\_lum\_0.5-10\_er+ & BXA Power-law fit Luminosity upper uncertainty in the 0.5-10 keV band (\ergs). \\
fit1\_pl\_logrelprob & $\log_{10}$ of goodness probability of the power-law model (normalized relative to the best model, which will have a probability of 1). \\
fit2\_apec\_nh & APEC fit N$_\mathrm{H}$ (cm$^{-2}$). \\
fit2\_apec\_nh\_er- & BXA APEC fit lower uncertainty (cm$^{-2}$). \\
fit2\_apec\_nh\_er+ & BXA APEC fit upper uncertainty (cm$^{-2}$). \\
fit2\_apec\_kT & BXA APEC temperature (keV). \\
fit2\_apec\_kT\_er- & BXA APEC temperature lower uncertainty (keV). \\
fit2\_apec\_kT\_er+ & BXA APEC temperature upper uncertainty (keV). \\
fit2\_apec\_flx\_0.5-10 & BXA APEC fit unabsorbed flux in the 0.5-10 keV band (\ergcms). \\
fit2\_apec\_flx\_0.5-10\_er- & BXA APEC fit unabsorbed flux lower uncertainty in the 0.5-10 keV band (\ergcms). \\
fit2\_apec\_flx\_0.5-10\_er+ & BXA APEC fit unabsorbed flux upper uncertainty in the 0.5-10 keV band (\ergcms). \\
fit2\_apec\_lum\_0.5-10 & BXA APEC fit Luminosity in the 0.5-10 keV band (assuming cluster distance; \ergs). \\
fit2\_apec\_lum\_0.5-10\_er- & BXA APEC fit Luminosity lower uncertainty in the 0.5-10 keV band (\ergs). \\
fit2\_apec\_lum\_0.5-10\_er+ & BXA APEC fit Luminosity upper uncertainty in the 0.5-10 keV band (\ergs). \\
fit2\_apec\_logrelprob & $\log_{10}$ of goodness probability of the APEC model (normalized relative to the best model, which will have a probability of 1). \\
fit3\_bbodyrad\_nh & BXA Blackbody fit N$_\mathrm{H}$ (cm$^{-2}$). \\
fit3\_bbodyrad\_nh\_er- & BXA Blackbody fit lower uncertainty (cm$^{-2}$). \\
fit3\_bbodyrad\_nh\_er+ & BXA Blackbody fit upper uncertainty (cm$^{-2}$). \\
fit3\_bbodyrad\_kT & BXA Blackbody temperature (keV). \\
fit3\_bbodyrad\_kT\_er- & BXA Blackbody temperature lower uncertainty (keV). \\
fit3\_bbodyrad\_kT\_er+ & BXA Blackbody temperature upper uncertainty (keV). \\
fit3\_bbodyrad\_flx\_0.5-10 & BXA Blackbody fit unabsorbed flux in the 0.5-10 keV band (\ergcms). \\
fit3\_bbodyrad\_flx\_0.5-10\_er- & BXA Blackbody fit unabsorbed flux lower uncertainty in the 0.5-10 keV band (\ergcms). \\
fit3\_bbodyrad\_flx\_0.5-10\_er+ & BXA Blackbody fit unabsorbed flux upper uncertainty in the 0.5-10 keV band (\ergcms). \\
fit3\_bbodyrad\_lum\_0.5-10 & BXA Blackbody fit Luminosity in the 0.5-10 keV band (assuming cluster distance; \ergs). \\
fit3\_bbodyrad\_lum\_0.5-10\_er- & BXA Blackbody fit Luminosity lower uncertainty in the 0.5-10 keV band (\ergs). \\
fit3\_bbodyrad\_lum\_0.5-10\_er+ & BXA Blackbody fit Luminosity upper uncertainty in the 0.5-10 keV band (\ergs). \\
fit3\_bbodyrad\_logrelprob & $\log_{10}$ of goodness probability of the Blackbody model (normalized relative to the best model, which will have a probability of 1). \\
best\_fit & Model with highest probability. \\
N\_bg\_agn\_soft & Number of expected bkg sources with flux higher than the source in the 0.5-2 keV, within the half-light radius of the GC based on \citet{Mateos08}. \\
N\_bg\_agn\_hard & Number of expected bkg sources with flux higher than the source in the 2-10 keV, within the half-light radius of the GC based on \citet{Mateos08}. \\
dist\_from\_gc\_center & Source distance from the cluster center (arcmin). \\
gc\_name & Host GC name \\
gc\_distance & GC distance (kpc) \\
gc\_nh & Host cluster N$_\mathrm{H}$ (cm$^{-2}$) as estimated based on cluster E(B-V). This value is not used in spectral analyses in this work and is provided for comparison. \\
gc\_nh\_er & Uncertainty in the host cluster N$_\mathrm{H}$ (cm$^{-2}$),  considering the reported uncertainties on $E(B-V)$ in the Harris catalog, the reported uncertainty on the correlation slope by \citet{Bahramian15}, and assuming an uncertainty of 0.1 on $R_V$. \\
gc\_cr\_rad & Host cluster core radius (arcmin). \\
gc\_hl\_rad & Host cluster half-light radius (arcmin). \\
\enddata
\end{deluxetable*}
\end{longrotatetable}


\bibliography{full_bib}{}
\bibliographystyle{aasjournal}



\end{document}